\documentclass[11pt]{article}

\usepackage{enumitem}
\usepackage{comment}
\usepackage[english]{babel}
\usepackage{amsthm}
\usepackage[utf8]{inputenc}
\usepackage{xcolor}
\usepackage{amsmath}
\usepackage{physics}
\usepackage[margin=1in]{geometry}
\usepackage{titlesec}
\usepackage{natbib}
\usepackage{graphicx}
\usepackage{graphicx} 
\usepackage{float}

\newtheorem{corollary}{Corollary}

\newtheorem{proposition}{Proposition}

\newtheorem{example}{Example}

\usepackage{caption}
\usepackage{subcaption}
\usepackage{amsmath}
\usepackage{array}
\usepackage{graphicx}
\usepackage{placeins}

\usepackage{algorithm}
\usepackage{algpseudocode}

\usepackage{amsmath}

\usepackage{xcolor}
\usepackage{hyperref}
\hypersetup{%
  colorlinks=false,
  linkbordercolor=red
}

\usepackage{natbib}
\usepackage{graphicx}

\usepackage{lineno}

\title{
\textbf{A Population Sampling Framework for Claim Reserving in General Insurance}
}

\author{
\textbf{Sebastián Calcetero Vanegas*, Andrei L. Badescu, X. Sheldon Lin} \\
Department of Statistical Sciences\\
  University of Toronto\\
  Toronto, Ontario \\
  *\texttt{sebastian.calcetero@mail.utoronto.ca} \\
  }

\begin{document}

\maketitle

\begin{abstract} 
Claim reserving in insurance has been studied through two primary frameworks: the macro-level approach, which estimates reserves at an aggregate level (e.g., Chain-Ladder), and the micro-level approach, which estimates reserves at the individual claim level \cite{antonio2014micro}. These frameworks are based on fundamentally different theoretical foundations, creating a degree of incompatibility that limits the adoption of more flexible models. This paper introduces a unified statistical framework for claim reserving, grounded in population sampling theory. We show that macro- and micro-level models represent extreme yet natural cases of an augmented inverse probability weighting (AIPW) estimator. This formulation allows for a seamless integration of principles from both aggregate and individual models, enabling more accurate and flexible estimations. Moreover, this paper also addresses critical issues of sampling bias arising from partially observed claims data—an often overlooked challenge in insurance. By adapting advanced statistical methods from the sampling literature, such as double-robust estimators, weighted estimating equations, and synthetic data generation, we improve predictive accuracy and expand the tools available for actuaries. The framework is illustrated using Canadian auto insurance data, highlighting the practical benefits of the sampling-based methods.

\end{abstract}


\section{Introduction}

Accurate estimation of reserves of outstanding claims is fundamental to insurance operations, ensuring solvency and compliance with regulatory requirements. The reserving problem in insurance has been extensively studied in actuarial science literature and remains an active area of research due to challenges arising from the data. Outstanding claims reserves are typically divided into two categories: Reported But Not Settled (RBNS) reserves and Incurred But Not Reported (IBNR) reserves, each requiring distinct estimation methodologies. RBNS reserves generally constitute a larger portion of the total reserve but are often of less analytical interest since they can be reasonably estimated using expert judgment, e.g., case estimates, given that the severeness of the associated accidents is known. In contrast, IBNR reserves involve significantly greater uncertainty, as no information is available about the claims they encompass. This makes IBNR reserves more challenging to estimate and a primary focus for both practitioners and researchers. In this discussion, for simplicity, we will focus our analysis on the IBNR reserves. However, we note that the principles outlined here can, with minimal adjustment, be applied to RBNS reserves as well.

The insurance industry has approached claim reserving through two distinct frameworks: macro-level and micro-level methodologies, each with its own theoretical foundations and practical applications (see, e.g., \cite{taylor2019loss} for a review of reserving techniques). The macro-level, or aggregate-level, approach—exemplified by the Chain-Ladder (CL) method—estimates reserves at the portfolio level, relying on aggregated data for simplicity and ease of use. However, this approach overlooks individual claim behavior, which can result in inaccuracies. In contrast, the micro-level, or individual-level, approach, as described by \cite{antonio2014micro}, utilizes claim-specific data to provide more precise reserve estimates. This precision, however, comes with increased complexity and higher computational demands. Notably, according to a survey by \cite{astinreport}, individual reserving methods remain virtually absent in practice, with insurance companies worldwide predominantly relying on aggregate models.

Aggregate and individual models have traditionally been built on distinct statistical foundations due to their differing data sources, often resulting in incompatible reserve estimates. This dichotomy has constrained the flexibility and adaptability of reserving methods, impeding the development of models that integrate the strengths of both approaches and slowing the adoption of advanced techniques in the insurance industry. Several studies, including \cite{calcetero2023claim}, \cite{tee2022estimating}, \cite{wahl2019collective}, \cite{wuthrich2018neural}, \cite{hiabu2017relationship}, and \cite{charpentier2016macro}, have sought to address this issue by proposing methods to bridge the aggregate and individual approaches. However, these contributions are often narrowly focused on specific techniques and fail to offer a unified framework that naturally accommodates both micro and macro perspectives of the reserving process.

In response to this challenge, \cite{calcetero2023claim} and \cite{CalceteroVanegas2024} recently introduced an inverse probability weighting (IPW) estimator that integrates granular information into the CL method, providing an intermediate structure between aggregate and individual models. This approach was developed by conceptualizing the reserving problem as a population sampling issue, as discussed in \cite{arnab2017survey}. Population sampling involves selecting and analyzing data subsets to infer characteristics of the entire population, often using sampling designs optimized for the population's structure. In the context of reserving, this design is similar to Poisson sampling, where each claim is either included or excluded from the sample (i.e., reported or not) based on a Bernoulli-like trial, which depends on the reporting delay time. This process is driven by the nature of the claims, rather than being determined by the analyst, and resembles mechanisms studied in missing data, where only available records are analyzed (e.g., surveys with non-response), and in causal inference, where data is observable only for those who were treated.

While the approach by \cite{calcetero2023claim} introduces a novel perspective on claim reserving and expands the CL method, it offers limited exploration of how the population sampling perspective can be fully applied to claim reserving problems. Additionally, it does not adequately address how this perspective can be extended to individual-level methods, thereby providing a comprehensive framework for understanding both micro and macro models. Specifically, \cite{calcetero2023claim} focused on the IPW estimator within the so called \emph{design-based} approach to sampling, which does not consider a probabilistic model for the data-generating process. Instead, it depends solely on the randomness of the sampling process, i.e., reporting delays. However, this approach has limitations, as it cannot incorporate distributions or other modeling insights for the claims data, e.g., heavy tails, that are often crucial and widely used in the reserving literature.

Along these lines, this paper addresses the gaps in \cite{calcetero2023claim} related to the population sampling perspective of the reserving problem, extending it to the \emph{model-based} approach to sampling, in which distributional models for the claims data are also incorporated. This extension is achieved through augmented inverse probability weighting (AIPW) estimators, which build on the IPW framework by integrating a model that includes distributional assumptions in the predictions. We demonstrate that the AIPW estimator provides a unified statistical framework for claim reserving, integrating both aggregate and individual-level approaches. Specifically, it incorporates models such as the Chain-Ladder and Bornhuetter-Ferguson models at the aggregate level, alongside individual models where the severity model satisfies the balance property. This creates a seamless transition between the macro and micro approaches. Thus, the AIPW estimator serves as a natural bridge between aggregate and individual reserving models, enabling the simultaneous leveraging of their respective strengths.

Moreover, this paper explores how the population sampling perspective broadens our understanding of the reserving process. In particular, it highlights issues related to sample bias arising from partially observed data—one of the main motivations for using AIPW estimators. Indeed, the sampling bias problem stems from the fact that the behavior of reported claims may differ from that of not reported claims. As a result, inferences based solely on reported claims data may not generalize to not reported claims, leading to poor predictions. This issue is common in insurance data but is often overlooked by actuaries, e.g.,  \cite{actuarial_data_bias_2023}. To address this, we adapt practical statistical approaches from the population sampling literature that can mitigate sampling bias in insurance reserving, thereby improving reserve predictions. In particular, we illustrate methods such as double-robust estimators, weighted estimating equations for model fitting, and synthetic data generation. These techniques not only enhance the precision and robustness of reserve estimates but also introduce statistical methods into the reserving process, which have not yet been used in the field. Our goal is to illustrate the application of the class of sampling-based methods and highlight their benefits for reserving, rather than showcasing the best configuration among them.

This paper is organized as follows: Section \ref{SampFrame} introduces the claim reserving problem, defines our notation, and revisits the population sampling perspective. Section \ref{bridgap} presents the AIPW framework as a unifying statistical tool for both aggregate and individual reserving methods. Section \ref{lessons} explores additional applications of sampling methods in reserving. Section \ref{numerical} demonstrates how these sampling methods can be applied to IBNR reserves, using a real Canadian auto insurance data to illustrate their practical applications. Finally, Section \ref{conclu} concludes with a discussion of the implications and directions for future research.

\section{A population sampling framework for claim reserving}
\label{SampFrame}

Suppose an insurance company is evaluating its total liabilities from claims with accident times between $t=0$ and $t=\tau$, where $\tau$ is the valuation time set by the actuary. In general insurance, there is often a delay between an accident occurrence and its reporting time to the company. Moreover, the situation becomes more complex due to payment delays. Reported claims are often paid in several installments as the accident's impact evolves. At valuation time $\tau$, the insurer knows only the claims that have been reported and the partial amount paid to each one. Therefore the insurer aims to estimate the total amount for not reported claims and the remaining payments for reported claims to calculate the reserve for outstanding claims.

While reported but not settled claims are of interest, insurance companies have more information regarding these as they know the graveness of the accident and can therefore provide an expert estimate of the liability, e.g. case reserves. Along those lines, we will focus on the liability of incurred but not reported claims, which are the claims the insurance company has no information. To address this, we will work with incurred losses rather than paid losses. Incurred losses consider the case reserve per claim, which represents the amount set aside by insurers for estimated unpaid costs of reported claims, based on expert judgment, current payments, and potential future developments. The sum of the case reserve and current payments forms the incurred losses, the best estimate of the total claim cost, which can be used as a reliable proxy for the value of a claim once is settled. From now on, we will work mainly with incurred losses thus avoiding the issue associated with the not settled claims. Similarly, we will refer to the incurred losses when mentioning the terms severity or claim amounts. This approach is common in reserving practices, where actuaries typically use incurred loss triangles rather than paid loss triangles to estimate IBNR reserves, and also in the literature, e.g.,  \cite{CalceteroVanegas2024}, \cite{fung2021new}.

Along those lines, let us describe the claim process as follows:

\begin{itemize}
\item $N(\tau)$: Number of claims with accident times before $\tau$. Note that this is partially unobserved before $\tau$.
\item $Y_{i}$: Sequence of incurred losses per claim, where $i = 1, \ldots, N(\tau)$.
\item $T_i$: Sequence of accident times for each claim, where $i = 1, \ldots, N(\tau)$.
\item $R_i$: Sequence of associated reporting times, where $i = 1, \ldots, N(\tau)$.
\item $U_i = R_i - T_i$: Sequence of reporting delay times for each claim, where $i = 1, \ldots, N(\tau)$.
\item $\boldsymbol{x}_i$: Sequence of attributes related to the accident, claim type, policyholder, or payment characteristics, where $i = 1, \ldots, N(\tau)$.
\item $N^{R}(\tau)$: Number of claims reported by $\tau$, and $N^{IBNR}(\tau)$: Number of incurred but not reported claims.
\end{itemize}

The total liability for accidents occurring before $\tau$, denoted $L(\tau)$, is:
$$
L(\tau) = \sum_{i=1}^{N(\tau)} Y_{i}.
$$
The portion of liability known to the company by $\tau$ (from reported claims), denoted $L^R(\tau)$, is:
$$
L^R(\tau) = \sum_{i=1}^{N^{R}(\tau)} Y_{i}.
$$
One is interested in estimating the liability for not reported claims, denoted $L^{IBNR}(\tau)$, which is:
$$
L^{IBNR}(\tau) = L(\tau) - L^R(\tau)=\sum_{i=1}^{N^{IBNR}(\tau) } Y_i .
$$

In reserving modeling, there are mainly two approaches explored in the literature in order to have an estimation of the outstanding claims. On one hand, there is the macro-level approach, with the Chain-Ladder (CL) as example, which works with aggregate data and provide estimates without explicitly imposing a probability model for the claim data (e.g. a frequency-severity model). These methods have been seen as ad-hoc procedures lacking statistical justification, and ignore the granular information from individual claims. On the other hand, there is the micro-level approach in which a model is used for claims and their attributes using probability distributions for the frequency, severity and reporting delays. As such, an actuary can derive a distributional model for the IBNR claims, and therefore use it to produce an estimation of the outstanding claims. Nevertheless, the reserving theory has been split mostly in these two, almost mutually exclusive branches, where no single framework provides a statistically sound umbrella for both approaches.

In response, a previous paper \cite{calcetero2023claim} illustrated a statistically justified methodology on which it is possible to introduce individual information on the CL method, producing therefore a hybrid aggregate and individual reserving method. As such, it aims to be a starting point to bridge the gap between the micro and macro model. This was achieved by a novel view of the reserving problem as a population sampling method as we review next.

\subsection{Claim reserving as a population sampling problem}
\label{Sampling_section}

Claim reserving can be seen as situation in which we want to estimate a population total based solely on a sample. In this context, all $N(\tau)$ claims are treated as the population, with the $N^{R}(\tau)$ claims reported by the valuation date forming the sample. Here one view this as \emph{Poisson sampling} without replacement, where each claim is a Bernoulli trial, either succeeding (reported by the valuation time) or failing. This is a similar sampling design associated to \emph{missing data} mechanisms. Unlike typical population sampling setups, the sample in reserving is given, rather than selected by an analyst. 

Along those lines, the statistical framework proceeds as follows: a membership indicator variable taking the value 1 if a claim is reported, and 0 otherwise can be defined as $\mathbf{1}_i(\tau) = \mathbf{1}_{ \{ R_i \le \tau \} }$, with the corresponding inclusion probabilities
$$
\pi_i(\tau) = P( R_i \le \tau \mid \boldsymbol{x}_i, T_i, Y_i) = P( U_i \le \tau - T_i \mid \boldsymbol{x}_i, T_i, Y_i).
$$

These probabilities, which depend on the valuation time and vary by claim due to differences in accident types and risk profiles, are unknown and must be estimated from the data. Estimation can be done using techniques such as logistic regression or survival analysis, and is often part of the estimation of reserving models themselves. Indeed, since reporting delay time is a time-to-event random variable with right truncation, survival modeling approaches, such as Cox regression, e.g. \cite{Vakulenko2019}, \cite{george2014survival}, can be used to estimate the cumulative distribution function, from which the inclusion probabilities can be derived. Proper estimation of these probabilities is crucial from the population sampling perspective as their the main input for all the methods that follow. This procedure will be further illustrated in the numerical study. From now on, we will use the notation $\hat{\pi}_i(\tau)$ to denote the estimated inclusion probabilities based on the selected model..

\subsection*{The IPW estimator}
Based on such a sampling design, \cite{calcetero2023claim} introduced the unbiased inverse probability weighting (IPW) estimator for the total liability, which scales the reported claims amounts by a factor of \(\frac{1}{\hat{\pi}_i(\tau)}\) using the expression:
$$
\sum_{i=1}^{N^R(\tau)} \frac{Y_i}{\hat{\pi}_i(\tau)}. 
$$

Consequently, an unbiased IPW estimator for the IBNR reserve,  which utilizes the odds ratio \(\frac{1-\hat{\pi}_i(\tau)}{\hat{\pi}_i(\tau)}\) that represent the frequency ratio of not reported to reported claims with similar characteristics, via the expression: 
\begin{equation}
\label{IPW_Y1}
\sum_{i=1}^{N^R(\tau)} \frac{1-\hat{\pi}_i(\tau)}{\hat{\pi}_i(\tau)} Y_i.
\end{equation}

Although this estimator may not seem intuitive at first glance, it is a natural choice from the perspective of claim reserving as motivated below. The following proposition illustrates how the IPW principle is an ingrained component in the estimation of reserves for not reported claims using reported claims as a starting point. Nevertheless, recall that the IPW principle arises on its own from the population sampling literature, rather than from the claim reserving problem. This proposition serves merely as an illustration and provides an initial result on how population sampling techniques, such as IPW, connect to claim reserving and contribute to a reserving framework.

\begin{proposition}
\label{IPWres}

Let \( \{N(t; \tau), t \leq \tau\} \) be a counting process representing the number of claims occurring before the valuation date \( \tau \), with an expected count function given by  \(
\Lambda(t) = E(N(t; \tau))\). Each claim is associated with time-dependent marks \( (T_i, Y_i, \mathbf{1}_i(\tau)) \). Suppose that the marks  \( Y_i \) and \( \mathbf{1}_i(\tau) \) are conditionally uncorrelated given \( T_i \), and have the following conditional expectations  $\mu(t)=E(Y\vert T=t)$ and $\pi(\tau-t)=E( \mathbf{1}(\tau) \vert T=t)$.

Let the corresponding compound processes \( \{L^R(t; \tau), t \leq \tau\} \) as the thinned process representing the liability of reported claims, i.e., $L^R(t; \tau) = \sum_{i=1}^{N(t; \tau)} Y_i \mathbf{1}_i(\tau) $; and  \( \{L^{IBNR}(t; \tau), t \leq \tau\} \) as the thinned process representing the liability of incurred but not reported (IBNR) claims, i.e.,  $L^{IBNR}(t; \tau) = \sum_{i=1}^{N(t; \tau)} Y_i (1 - \mathbf{1}_i(\tau))$.

Then the following relationship holds between the expected values:
\begin{equation}
\label{IPWform}
E(L^{IBNR}(t;\tau)) = \int_0^{t} \frac{1-\pi(\tau-s)}{\pi(\tau-s)}dE(L^R(s;\tau))
\end{equation}
where this is a Riemann–Stieltjes integral with respect to  $E(L^R(t;\tau))$, as a function of $t$.

\end{proposition}

\begin{proof}
A direct application of iterated expectation and Campbell's theorem leads to:
$$
E(L^{R}(t;\tau)) = E \left( \sum_{i=1}^{N^R(t;\tau)} Y_i \mathbf{1}_i(\tau)  \right) = E \left( \sum_{i=1}^{N^R(t;\tau)} \mu(T_i) \pi(\tau-T_i) \right)  = \int_0^{t} \mu(s)\pi(\tau-s)d\Lambda(s) 
$$
and similarly, $E(L^{IBNR}(t;\tau)) = \int_0^{t} \mu(s)(1-\pi(\tau-s))d\Lambda(s)$. Equivalently, in differential form we obtain:
$$
dE(L^{R}(t;\tau)) = \mu(t)\pi(\tau-t)d\Lambda(t)  \ \ \ \textrm{   and   } \ \ \ dE(L^{IBNR}(t;\tau)) =\mu(t)(1-\pi(\tau-t))d\Lambda(t),
$$
and after a simple rearrangement, we obtain
$$
dE(L^{IBNR}(t;\tau)) = \frac{1-\pi(\tau-t)}{\pi(\tau-t)}dE(L^R(t;\tau)),
$$
which leads to the integral form shown above.
\end{proof}
Proposition \ref{IPWres} provides an interpretation of the odds ratio of probabilities as a change-of-measure mechanism, similar to a Radon-Nikodym derivative. Indeed, the IPW principle acts as a weighting function within an integral that transforms expectations of aggregate reported claims into expectations of aggregate not reported claims. This idea can be used to show results associated to the IPW principle, as we will do in the next sections. Moreover,  this proposition  can be used to motivate the IPW estimator in Equation (\ref{IPW_Y1}) as an empirical version of Equation \ref{IPWform} where we use the 
 observed path of $L^R(t;\tau)$ as integrator. This assertion can be evidenced as an interim step  in Corollary \ref{unbiased}. 

\begin{corollary}
\label{unbiased}
The IPW estimator for the IBNR reserve is unbiased.
\end{corollary}
\begin{proof}
Interchanging the integral and expectation on the right-hand side of Equation(\ref{IPWform}) we obtain
$$
E(L^{IBNR}(\tau)) =E(L^{IBNR}(\tau;\tau))  = E \left( \int_0^{\tau} \frac{1-\pi(\tau-s)}{\pi(\tau-s)}dL^R(s;\tau) \right) = E\left( \sum_{i=1}^{N^R(\tau)} \frac{1-\pi_i(\tau)}{\pi_i(\tau)} Y_i\right).
$$
The integral simplifies to a sum as $L^R(t;\tau)$ is a step function. Note this proof differs from that from the sampling literature. The latter is more general as does not require a probability model.
\end{proof}

One of the main attributes of the IPW estimator is that it achieves an estimation in a frequency-severity distribution-free manner, similar to the CL method. In fact, \cite{calcetero2023claim} demonstrated that the CL method can be viewed as an IPW estimator where inclusion probabilities are approximated using empirical proportions. By employing a predictive model to estimate inclusion probabilities with granular information, the basic CL method can be extended to incorporate more granular data. We will revisit these results in the next section.

While such a result is remarkable, the discussion on such paper did not fully explore the broader context of the population sampling literature. Specifically, the IPW approach was examined under the \emph{design-based} framework, which relies solely on the randomness of the sampling mechanism (i.e., whether a claim is or not reported) for inference, without accounting for the probabilistic model underlying claim generation process.  In contrast, the population sampling literature encompasses more general frameworks, such as the \emph{model-based} approach. This approach incorporates a probabilistic model for claim generation alongside the randomness of the sampling design, thereby leveraging additional information for more accurate modeling and predictions. In this setup, the goal is to combine this model with the IPW  to estimate population totals more effectively.  

The broader perspective of the \emph{model-based} approach enriches our understanding of reserving and enables more robust methods for prediction. As we will demonstrate in the next sections, sampling-based methods under the \emph{model-based} approach provide a statistically sound foundation for reserving practices, offering valuable insights for improved predictions.


\section{Bridging the gap between aggregate and individual reserving using AIPW estimators}
\label{bridgap}

This section introduces new augmented inverse probability weighting (AIPW) estimators for the reserve, under the model based approach. Our goals with such estimators are 1) to extend and improve the usual IPW estimator for the reserve with a more robust alternative and 2) to provide a unified statistical framework for macro and micro claim reserving methods. Historically, many aggregate reserving methods were devised as ad-hoc procedures or derived artificially in the stochastic claim reserving literature under non-intuitive assumptions about the claim generation mechanism. In contrast, our framework shows a construction of these methods at an individual level, which provides a more natural interpretation, and an easy pathway to integrate with individual methods. This insight positions the theory of AIPW estimators and population sampling as a cohesive foundation for claim reserving, offering a fresh perspective on these methodologies.

We begin by introducing AIPW estimators as a natural extension of IPW estimators on which a model for the claim data generation is at a play, and then discuss how aggregate and individual reserving approaches are special cases of AIPW estimators, under specific conditions, and therefore bridging the gap for these methodologies.

\subsection{Augmented inverse probability weighting estimators for the reserve}

Consider a reserving model already selected and fitted to the claim data, which is based on some distributional assumptions, e.g. an individual reserving model or a loss-ratio-based model, etc. This model serves as the starting point for further predictions. Typically, such a model—built using frequency, severity, and reporting delay time models—provides estimates of the total number of claims \(\hat{N}(\tau)\), the total number of non-reported claims \(\hat{N}^{IBNR}(\tau)\), and the amounts for each claim, denoted by \(\hat{Y}_i\). The total reserve for unobserved claims is then estimated as \(\sum_{i=1}^{\hat{N}^{IBNR}(\tau)} \hat{Y}_i\). Here, the term "model" is used broadly, as these estimates may derive from qualitative or quantitative sources, not necessarily from a statistical model. For instance, the estimates could be informed by expert opinion, such as for case estimates, or by predictive models, as in \cite{antonio2014micro}.

As inspired in the literature of population sampling, e.g., \cite{sarndal2003model}, we propose to extend the IPW estimator to a \emph{ model assisted} version via an \emph{augmented inverse probability weighting} (AIPW) estimator (also known as the difference estimator) given by Equation (\ref{AIPW_ult}).

\begin{equation}
\hat{L} (\tau) = \sum_{i=1}^{\hat{N}(\tau)} \hat{Y}_i + \sum_{i=1}^{N^{R}(\tau)} \frac{Y_i-\hat{Y}_i}{\hat{\pi}_i(\tau)},
\label{AIPW_ult}
\end{equation}
where the first term represents the estimation of the population total based solely on the reserving model through the  estimates $\hat{Y}_i$, and the second term is a scaled version of the prediction errors of the model ($Y_i - \hat{Y}_i$) in the sample. The motivation behind considering this type of estimator is based on applying the IPW principle, but on the errors of the model as in the following idea:

$$
L(\tau) = \sum_{i=1}^{N(\tau)} Y_i = \sum_{i=1}^{N(\tau)} \hat{Y}_i + \sum_{i=1}^{N(\tau)} (Y_i - \hat{Y}_i)  \approx \sum_{i=1}^{\hat{N}(\tau)} \hat{Y}_i + \sum_{i=1}^{N^{R}(\tau)} \frac{Y_i-\hat{Y}_i}{\hat{\pi}_i(\tau)}. 
$$
The intuition here is that the errors  $Y_i-\hat{Y}_i$ potentially have less variability than the outcome $Y_i$, and therefore the associated AIPW estimator would have a lower variance and would lead to better predictions, while also being an unbiased estimator as we show below.

Along those lines, we can derive an unbiased AIPW estimator of the outstanding claims as
 \begin{align} 
 \label{AIPW_Y1}
\hat{L}^{IBNR}(\tau) = \hat{L}(\tau)-L^R(\tau) &= \sum_{i=1}^{\hat{N}(\tau)} \hat{Y}_i + \sum_{i=1}^{N^{R}(\tau)} \frac{Y_i-\hat{Y}_i}{\hat{\pi}_i(\tau)} - \sum_{i=1}^{N^{R}(\tau)} Y_i \\ \label{AIPW_Y2}
& =  \sum_{i=1}^{N^{R}(\tau)} \frac{1-\hat{\pi}_i(\tau)}{\hat{\pi}_i(\tau)} Y_i+ \left( \sum_{i=1}^{\hat{N}^{IBNR}(\tau)} \hat{Y}_i -\sum_{i=1}^{N^{R}(\tau)} \frac{1-\hat{\pi}_i(\tau)}{\hat{\pi}_i(\tau)}\hat{Y}_i \right) \\ \label{AIPW_Y3}
& =  \sum_{i=1}^{\hat{N}^{IBNR}(\tau)} \hat{Y}_i + \sum_{i=1}^{N^{R}(\tau)} \frac{1-\hat{\pi}_i(\tau)}{\hat{\pi}_i(\tau)} (Y_i-\hat{Y}_i), 
\end{align}

\noindent in which we use the decomposition $\sum_{i=1}^{\hat{N}^{}(\tau)} \hat{Y}_i = \sum_{i=1}^{N^{R}(\tau)} \hat{Y}_i+\sum_{i=1}^{\hat{N}^{IBNR}(\tau)}\hat{Y}_i$ (abusing the notation of the indices) in the last equality. Note that the regular IPW estimator is recovered by ignoring the terms $\hat{Y}_i$ from the model assisting.

The AIPW estimator acts as a correction mechanism. The first term in Equations (\ref{AIPW_Y2}) and  (\ref{AIPW_Y3}) represent the IPW estimator and the micro level estimator for the IBNR reserve respectively, while the associated second term addresses errors in the estimators. Specifically, in Equation (\ref{AIPW_Y3}), the second term corrects the micro-level prediction by adding the mismatch between such predictions and the observed reported claims, i.e., $Y_i-\hat{Y}_i$, after being stretched by the inclusion probabilities odds ratio. In Equation (\ref{AIPW_Y2}), the second term corrects the usual IPW estimator by comparing the predicted and weighted-scaled estimates for not reported claims, focusing on discrepancies caused by the IPW scaling. Thus the reserving model and the IPW estimator correct each other, providing robustness against misspecification. 

This interplay between the reserving model and the IPW estimator ensures a desirable property, widely studied in statistics,  known as double robustness: if either a) the estimated inclusion probabilities are correctly specified (i.e., the estimated probabilities match the true inclusion probabilities) or b) the prediction of the reserving model of the population total is unbiased, the AIPW estimator provides an unbiased reserve estimate (see, e.g., \cite{seaman2018introduction}).   As most of the literature presents this property in the context of causal inference and setups where the population size is deterministic,  we establish and demonstrate a version within our framework for completeness in Proposition \ref{unbias} below. 

\begin{proposition}[Double robustness of the AIPW]
\label{unbias}
Under the notation introduced so far, assume also that a) the frequency model is conditionally unbiased given the vector of attributes, i.e., \( E \left(  \hat{N}(\tau) \mid \{\mathbf{x}_i\}_{i=1}^{N^R(\tau)} \right)  = E \left(  N(\tau) \mid \{\mathbf{x}_i\}_{i=1}^{N^R(\tau)} \right)  \), and  b) that frequency and severity are conditionally independent, given the vector of attributes.

Therefore the AIPW estimators in Equations (\ref{AIPW_ult}) and (\ref{AIPW_Y3})  are unbiased, i.e., \( E \left( \hat{L}(\tau) \mid \{\mathbf{x}_i\}_{i=1}^{N^R(\tau)} \right) = E \left(  L(\tau) \mid \{\mathbf{x}_i\}_{i=1}^{N^R(\tau)} \right)  \) and \( E \left( \hat{L}^{IBNR}(\tau) \mid \{\mathbf{x}_i\}_{i=1}^{N^R(\tau)} \right)  = E \left(  L^{IBNR}(\tau) \mid \{\mathbf{x}_i\}_{i=1}^{N^R(\tau)} \right) \), provided at least one of the following holds:

\begin{enumerate}
\item The inclusion probabilities are correctly specified, that is, $\hat{\pi}_i(\tau)=\pi_i(\tau) = E( \mathbf{1}_i(\tau) \mid Y_i, T_i,\mathbf{x}_i ) $, where recall $\mathbf{1}_i(\tau)$ is the membership indicator random variable.

\item The severity model is conditionally unbiased, that is,  $E(\hat{Y}_i \mid  \mathbf{1}_i(\tau), T_i, \mathbf{x}_i )=E(Y_i \mid  \mathbf{1}_i(\tau), T_i, \mathbf{x}_i )$.

\end{enumerate}

\end{proposition}

\begin{proof}
We prove the assertion only for $\hat{L}(\tau)$ as it would follow immediately for $\hat{L}^{IBNR}(\tau)= \hat{L}^{}(\tau)-L^{R}(\tau)$ by linearity of the expectation. To do so we consider the two cases separately:

\begin{enumerate}
\item Suppose the inclusion possibilities are correctly specified, as $\hat{\pi}_i(\tau)=\pi_i(\tau)=E( \mathbf{1}_i(\tau) \mid Y_i, T_i,\mathbf{x}_i )$. Then we have, 
\begin{align*}
E \left( \hat{L}(\tau) \mid \{\mathbf{x}_i\}_{i=1}^{N^R(\tau)} \right) & =  E \left( \sum_{i=1}^{\hat{N}(\tau)} \hat{Y}_i + \sum_{i=1}^{N^{R}(\tau)} \frac{Y_i-\hat{Y}_i}{\hat{\pi}_i(\tau)} \mid \{\mathbf{x}_i\}_{i=1}^{N^R(\tau)} \right) \\
& =  E \left( \sum_{i=1}^{N(\tau)} \hat{Y}_i \mid \{\mathbf{x}_i\}_{i=1}^{N^R(\tau)} \right)  + E \left(  \sum_{i=1}^{N(\tau)} \frac{Y_i-\hat{Y}_i}{\pi_i(\tau)} \mathbf{1}_i(\tau)\mid \{\mathbf{x}_i\}_{i=1}^{N^R(\tau)} \right) \\
& = E \left( \sum_{i=1}^{N(\tau)} \hat{Y}_i \mid \{\mathbf{x}_i\}_{i=1}^{N^R(\tau)} \right) + E \left(  \sum_{i=1}^{N(\tau)} \frac{Y_i-\hat{Y}_i}{\pi_i(\tau)} E(\mathbf{1}_i(\tau)\mid Y_i, T_i,\mathbf{x}_i ) \mid \{\mathbf{x}_i\}_{i=1}^{N^R(\tau)} \right)  \\
& =  E \left( \sum_{i=1}^{N(\tau)} \hat{Y}_i \mid \{\mathbf{x}_i\}_{i=1}^{N^R(\tau)} \right)  + E \left(  \sum_{i=1}^{N(\tau)} Y_i-\hat{Y}_i \mid \{\mathbf{x}_i\}_{i=1}^{N^R(\tau)} \right)\\
& = E \left(  \sum_{i=1}^{N(\tau)} Y_i \mid \{\mathbf{x}_i\}_{i=1}^{N^R(\tau)} \right) = E \left( L(\tau) \mid \{\mathbf{x}_i\}_{i=1}^{N^R(\tau)} \right),
\end{align*}
where we use the assumption that the frequency model is unbiased and the estimated inclusion probability are correctly specified to remove the associated hats in the second equality, apply the iterated law of expectation (conditioning on claim sizes and other attributes) in the third, and cancel out common terms in the fourth. Notably, no assumptions were made about \(\hat{Y}_i\).

\item Suppose that the claim severity model is unbiased, so that $E(\hat{Y}_i \mid \mathbf{1}_i(\tau), T_i, \mathbf{x}_i )=E(Y_i \mid \mathbf{1}_i(\tau), T_i, \mathbf{x}_i )$. Then we have, 
\begin{align*}
E \left( \hat{L}(\tau) \mid \{\mathbf{x}_i\}_{i=1}^{N^R(\tau)} \right) & =  E \left( \sum_{i=1}^{\hat{N}(\tau)} \hat{Y}_i + \sum_{i=1}^{N^{R}(\tau)} \frac{Y_i-\hat{Y}_i}{\hat{\pi}_i(\tau)} \mid \{\mathbf{x}_i\}_{i=1}^{N^R(\tau)} \right) \\
& =  E \left( \sum_{i=1}^{N(\tau)} Y_i \mid \{\mathbf{x}_i\}_{i=1}^{N^R(\tau)} \right) + E \left(  \sum_{i=1}^{N(\tau)} \frac{Y_i-\hat{Y}_i}{\hat{\pi}_i(\tau)} \mathbf{1}_i(\tau) \mid \{\mathbf{x}_i\}_{i=1}^{N^R(\tau)} \right) \\
& = E \left( \sum_{i=1}^{N(\tau)} Y_i \mid \{\mathbf{x}_i\}_{i=1}^{N^R(\tau)} \right) + E \left(  \sum_{i=1}^{N(\tau)} \frac{E(Y_i-\hat{Y}_i \mid \mathbf{1}_i(\tau), T_i, \mathbf{x}_i )}{\hat{\pi}_i(\tau)}  \mathbf{1}_i(\tau)\mid \{\mathbf{x}_i\}_{i=1}^{N^R(\tau)} \right)  \\
& =  E \left( \sum_{i=1}^{N(\tau)} Y_i \mid \{\mathbf{x}_i\}_{i=1}^{N^R(\tau)} \right) +0 = E \left( L(\tau) \mid \{\mathbf{x}_i\}_{i=1}^{N^R(\tau)} \right),
\end{align*}
where we assume the frequency is unbiased to remove the associated hats in the second equality, apply the iterated law of expectation (conditioning on the indicators and other attributes) in the third, and cancel the second term in the fourth, based on the assumption of the severity model being unbiased. Notably, no assumptions were made about \(\hat{\pi}_i(\tau)\).

\end{enumerate}
Thus, the AIPW is a doubly robust estimator of the reserve, requiring only one of the models for the claim severity model or the inclusion probabilities to be correctly specified to have unbiasedness.
\end{proof}

The double-robustness property makes inference based on AIPW estimators more reliable than using a standard IPW estimator or a reserving model alone. The predictive performance of the AIPW estimator falls between that of the reserving model and the IPW estimator. A key takeaway from the sampling approach is that reserving models can benefit from the error correction provided by AIPW estimators, yielding more accurate estimates, as illustrated in the numerical study. However, this is more of a theoretical advantage than a strictly practical one. In reality, all models are misspecified to some extent, and inaccuracies in the reserving model or inclusion probabilities—often due to oversimplified assumptions—may limit the full realization of the double-robustness property. Nevertheless, AIPW estimators typically provide competitive estimates even under these conditions (e.g., \cite{waernbaum2023model}, \cite{kang2007demystifying}).


In insurance reserving, many methods implicitly incorporate principles related to the AIPW framework to improve accuracy. These methods are often conceived for practical purposes rather than being purely statistically motivated. In what follows, we demonstrate how the AIPW framework provides a statistical foundation for understanding various reserving methods.

\subsection{Aggregate reserving models as AIPW estimators}

Aggregate or macro-level reserving methods, such as the CL method and its extensions, rely on aggregate data. A key contribution of \cite{calcetero2023claim} was showing that the CL method is a particular case of the IPW estimator under certain homogeneity assumptions regarding the portfolio and inclusion probabilities. Specifically, for claims from a fixed accident year, say the $k$-th year, the CL estimator for ultimate claims can be expressed as:

\[
L^{CL}(\tau) = \frac{L^R(\tau)}{\pi^{CL}(\tau)},
\]
where \(\pi^{CL}(\tau)\) represents the implied inclusion probability, derived as the inverse of the CL development-to-ultimate factor, i.e., \(\pi^{CL}(\tau) = 1/f_k(\tau)\). Note that this idea applies similarly to the other accident years, and so it generalizes immediately. Along those lines, this formulation shows that the CL method is a particular case of the IPW estimator and, therefore, also an AIPW estimator,  albeit without incorporating an assisting model. 

Within the AIPW framework, this connection extends to other aggregate reserving methods. For example, as demonstrated in Proposition \ref{BF_example}, methods in the class of linear credibility models can be shown to be AIPW estimators. This class includes the  Bühlmann-Straub model, the Benktander-Hovinen method, the Cape-Cod method and other Bayesian approaches (see \cite{wuthrich2008stochastic}). This interpretation grants these reserving methods some of the theoretical properties of the AIPW framework, and extends the view of the population sampling beyond the CL method.

\begin{proposition}
\label{BF_example}
Consider an estimation $\hat{L}_{Cred}(\tau)$ that combines expert opinion with the Chain Ladder method to estimate the ultimate claims using a linear credibility approach, i.e., for a fixed accident year the ultimate claims are estimated as, 
\[
\hat{L}_{Cred}(\tau) = Z \hat{L}_{CL}(\tau) + (1 - Z) \hat{L}_{E}(\tau),
\]
where \(\hat{L}_{CL}(\tau)\) denotes the Chain Ladder estimate of the ultimate claims,  \(\hat{L}_{E}(\tau)\) the expert-based estimate and  \( Z \) is a credibility factor satisfying \( 0 \leq Z \leq 1 \).  

Then $\hat{L}_{Cred}(\tau)$ is an AIPW estimator, where the model assisting is the expert opinion, and the augmentation term is built based on the CL and its implied inclusion probabilities.
\end{proposition}

\begin{proof}
Define $\sum_{i=1}^{\hat{N}^{IBNR}(\tau)} \hat{Y}_i:= \hat{L}_{E}(\tau)$ as the reserving model estimate, and $\hat{L}^R_{E}(\tau) = \hat{L}_{E}(\tau) \times  \pi^{CL} (\tau) $ as the expected value for the current paid amount by $\tau$, as based on the expert opinion. Defining $ \pi^{CL} (\tau) = \pi^{CL} (\tau) /Z$, the credibility-based estimator becomes:
$$
\hat{L}_{Cred}(\tau)=\hat{L}_{E}(\tau) +\frac{ L^R(\tau) - \hat{L}^R_{E}(\tau)  }{  \pi^{CL} (\tau)} .
$$

This estimator is in structure an AIPW estimator where the model assisting is the expert opinion, and the augmentation term is built based on credibility adjusted CL inclusion probabilities. Lastly, note that different accident years are treated analogously, so the total reserve for all accident years follows immediately by adding up the results.
\end{proof}

\subsection{Individual reserving models as AIPW estimators}
We now delve into individual or micro-level models in insurance, which work with granular data. Micro-level reserving models aim to characterize each component involved in reserve estimation through predictive modeling at an individual level. This includes modeling the number of claims per individual (e.g. using a Poisson process), the corresponding claim amounts (e.g. using a GLM), and their associated reporting delay times (e.g. using survival model). Models are typically selected from flexible classes from the machine learning literature, chosen in a data-driven manner to accurately capture the behavior observed in the data. Consequently, an actuary can derive a distributional model for claims, allowing the estimation of a point estimate $\hat{Y}_i$ for claim amounts, as well as estimates for associated delays and the total number of claims, and non-reported claims $\hat{N}^{IBNR}(\tau)$. We refer to \cite{antonio2014micro} for more details.

To show when a micro-level model can be regarded as a specific case of an AIPW estimator, we first recall the concept of the \emph{balance property} of the severity model, as discussed in \cite{wuthrich2020bias} and \cite{campo2023insurance}. In summary, a severity model satisfies the balance property when the average predicted claim sizes align with the average observed values. In terms of totals, this can be expressed as:
$$
\sum_{i=1}^{N^{R}(\tau)} w_i\hat{Y}_i = \sum_{i=1}^{N^{R}(\tau)} w_iY_{i} ,
$$
where the $w_i$ serve as exposure-related factors, which can take values of $w_i=1$ for single exposure, or some other values depending on the context. The balance property guarantees that aggregate estimations from a predictive severity model match the losses at the portfolio level, which is the liability of interest. The importance of this property has been highlighted in both ratemaking and reserving contexts extensively.  With this in mind we now show how micro-level models can be seen as AIPW estimators.

\begin{proposition}
Suppose a micro-level model satisfies a weighted balance property with weights $w_i=\frac{1-\hat{\pi}_i(\tau)}{\hat{\pi}_i(\tau)}$, then the micro-level prediction is an AIPW estimator.
\end{proposition}

\begin{proof}
It suffices to verify when the predictions of the micro-level model align with those of the corresponding AIPW estimator. This can be observed in Equation (\ref{AIPW_Y3}), where the difference lies in the second term, the augmentation term. By assumption the model satisfies the weighted balance property, that is:
$$
   \sum_{i=1}^{N^{R}(\tau)} \frac{1-\hat{\pi}_i(\tau)}{\hat{\pi}_i(\tau)} \hat{Y}_i = \sum_{i=1}^{N^{R}(\tau)} \frac{1-\hat{\pi}_i(\tau)}{\hat{\pi}_i(\tau)} Y_i \Leftrightarrow \sum_{i=1}^{N^{R}(\tau)} \frac{1-\hat{\pi}_i(\tau)}{\hat{\pi}_i(\tau)} (Y_i-\hat{Y}_i) = 0,
$$
Hence, the augmentation term is negligible and the AIPW estimation becomes equivalent to the micro-level model.

\end{proof}

Several actuarial models satisfy the balance property naturally, such as any Generalized Linear Model (GLM) with a canonical link (ee, e.g., \cite{wuthrich2023statistical} for further details).  In cases where the balance property is not inherently satisfied, actuaries often adjust the severity model post-estimation to enforce this alignment (e.g., \cite{crevecoeur2022hierarchical}, \cite{campo2023insurance}). Consequently, most micro-level reserving models in practice happen to satisfy the balance property, ensuring their predictions align with those derived from the AIPW estimators, as long as the proper weight is used.  The interpretation and significance of this particular weighting scheme will be elaborated upon in the next section. This connection offers a compelling and natural rationale for why achieving the balance property in a model is desirable: it aligns the reserving model with the general properties of the AIPW estimator. Nevertheless, there is no single method to impose the balance property, leading to variations in model predictions and the resulting properties.

\subsection{Bridging the gap and hybrid approaches}
\label{gap}
Since these aggregate and individual reserving models are AIPW estimators, the AIPW framework provides a statistically sound approach to reserve estimation, unifying individual and aggregate models while drawing from population sampling theory. This perspective offers a novel interpretation of reserving methods, complementing existing results in stochastic claim reserving. Although beyond the scope of this paper, several noteworthy implications follow. For instance, inference techniques developed for AIPW estimators can now be applied to reserving methods, particularly to aggregate models, where statistical inference has lacked rigor. Additionally, key properties such as unbiasedness, optimality, and other nonparametric guarantees of reserving methods could be directly derived from the established theory of AIPW estimators.

Moreover, alternative AIPW structures can balance the simplicity of the CL method with the complexity of an individual reserving model within a single estimation, leading to hybrid approaches. Example \ref{ML-CL} illustrates this idea, and we assess its performance in our numerical study. Other hybrid methods can also be explored depending on analytical constraints. For instance, a standard IPW estimator with a predictive model solely for the inclusion probabilities, incorporating individual information, could serve as a hybrid approach, generalizing the CL method by integrating individual-level data.

\begin{example}
\label{ML-CL}
A potential structure of interest integrates directly the CL method and a microlevel model into a single estimate. To see how this can be achieved, note that in Equation (\ref{AIPW_Y2}), the IPW component of the AIPW estimator reduces to the Chain-Ladder prediction if we set the inclusion probabilities as \(\hat{\pi}_i^{CL}(\tau) = 1/f_i(\tau)\), where \( f_i(\tau) \) is the development factor to ultimate from the CL method. The second term then acts as an error correction term based on the micro-level model.  

Under this approach, the resulting AIPW estimator takes the form  
\[
\sum_{i=1}^{\hat{N}^{IBNR}(\tau)} \hat{Y}_i+\sum_{i=1}^{N^{R}(\tau)} \frac{1-\pi^{CL}_{i}(\tau)}{\pi^{CL}_{i}(\tau)}(Y_{i}-\hat{Y}_i),
\]
providing a direct trade-off between the micro-level model and the CL estimator. Similarly to the double robustness property of AIPW estimators, this estimator shifts weight toward either the CL estimate or the micro-level estimate, depending on their relative accuracy as reflected in the error correction terms. This balance makes it a compelling structure for reserving, serving as a middle ground between the two approaches.  

Note that this estimator may not necessarily outperform an AIPW estimator with model-based inclusion probabilities. However, as model misspecification is common in practice, especially with the challenge of modeling reporting delays, this alternative AIPW estimator may still offer a balanced approach with competitive predictive performance.

\end{example}

Lastly, while AIPW estimators provide a robust approach to reserving by integrating macro- and micro-level methods, they represent just one tool within the broader population sampling framework. Other methodologies can further address the challenges in reserving. In the next section, we explore these complementary techniques and their applications.

\section{Further applications of population sampling in  reserving}
\label{lessons}

The previous sections demonstrated the versatility of AIPW estimators in reserve estimation. However, more methodologies from the sampling literature can also offer valuable tools for reserving. This section highlights key ideas from population sampling that can improve the claim reserving predictions, providing more flexible and robust estimates.

Before introducing the methods, we first discuss the sampling bias issues associated with the reserving problem more thoroughly so that we can view the situation in an even more general setting. With this notion introduced, we then discuss the overall methods from the sampling literature designed to address such issues, for which the AIPW is a particular case.

\subsection{The sampling issues in the reserving problem}
In general sampling scenarios, the representativeness of a sample depends on the sampling design. When the sample is not representative of the overall population, the behavior of the sampled data may significantly differ from that of the population, leading to biased inferences, analysis, model fitting, among others if this is not addressed. This issue is known as sampling bias in the population sampling literature, and similar fields.

In the specific context of reserving, the reporting mechanism divides claim data into two groups: reported and not reported claims. The observed data of reported claims $(T_i, Y_i, U_i)$ come from a truncated distribution, as only claims reported by the valuation time are available, while the rest are treated as missing. Mathematically, the observed data from reported claims are realizations of the conditional distribution: $$
f_{T,Y,U}(t,y,u \mid  U \le \tau-T ) = \frac{f_{T,Y,U}(t,y,u) \mathbf{1}_{ \{u \le \tau -t \}} }{ P( U \le \tau -T)}.
$$
As a result, the behavior of reported claims may differ from that of not reported claims, i.e., \(f_{T,Y,U}(t,y,u \mid U \leq \tau-T) \neq f_{T,Y,U}(t,y,u \mid U > \tau-T)\), particularly if significant dependencies exist among these variables, leading to sampling bias on the reported claims data. Empirical evidence supports this distinction in auto insurance, where reported claims are often more severe than not reported ones. 

Along those lines, when estimating reserves and fitting reserving models for components such as claim severity, accounting for conditional distributions is essential to address sampling bias. Unfortunately, this is often overlooked by assuming independence of the claim severity and the reporting delay, e.g., $f_Y(Y_{i} \mid U_i \le \tau-T_i) = f_Y(Y_{i} \mid U_i > \tau-T_i) = f_Y(Y_{i})$. This simplification facilitates model fitting but ignores potential discrepancies due to the conditional nature of the data. Imposing a dependence structure between incurred losses and reporting delays would require methods like copulas, which are challenging due to the right-truncation inherent in reporting delays. However, the assumption of independence between claim severity and reporting may oversimplify the problem and lead to poor performance of predictions. If significant dependencies exist, reported claims may differ significantly in severity from not-reported ones, and neglecting these differences can undermine reserve accuracy. A numerical study in Section \ref{numerical}, using real data, illustrates this behavior.

To improve predictions, it is crucial to address discrepancies between reported and not reported claims cost-effectively when estimating reserving model components. This can be achieved by adopting a population sampling perspective and leveraging established techniques from that field. These methods are designed to mitigate sampling bias and enhance predictive accuracy but remain underutilized in actuarial science and reserving, in particular. The following subsection demonstrates how these approaches can be integrated into reserving models.

\subsection{Population sampling methods to handle sampling bias}

Here, we present an overview of population sampling techniques used to address sampling bias in model predictions. For simplicity, we focus on claim severity, though these methods apply equally to all components of the problem. In this context, we observe data from $f(Y \mid U \le \tau-T)$ i.e. reported claims,  but our goal is to estimate quantities related to  not reported claims $f(Y \mid U > \tau-T)$ (e.g. the mean or the whole distribution itself ).

The population sampling techniques for addressing biases on a severity model can be categorized based on when they are applied during the estimation process of such a model. Using terminology borrowed from fairness and discrimination bias mitigation (e.g., \cite{charpentier2024insurance}), we informally classify these methods in order of their prevalence in the literature:

\begin{itemize} 
\item Post-Processing: Adjustments for the sampling design are applied to predictions after fitting the severity model.
\item In-Processing: Adjustments are incorporated into the model fitting process to account for the sampling design.
\item Pre-Processing: Modifications are made directly to the data before fitting the model to address the sampling design.
\end{itemize}

All techniques begin with estimating inclusion probabilities, which define the sampling mechanism in the data and serve as a fundamental input for population sampling methods. In reserving, this involves estimating the probability of a claim being reported or not reported, as detailed in Section \ref{Sampling_section}. These probabilities are typically derived as part of the reserving model using survival models, an approach we demonstrate in our numerical case study. We will now discuss several methods under these approaches in detail.

\subsection{Post-processing approach: AIPW and weighted balance property}
\label{dbp}

In this approach, the data is treated as if there was no sampling bias, and the severity model is fitted using the data from the reported claims under the independence assumption. The correction associated with the sampling bias is then applied after to the predictions made by this model.

The AIPW, as introduced in the previous section, falls into this category. The augmentation term serves as a post-processing correction to the overall prediction of the reserve to the given individual reserving model.   Specifically, this term can be interpreted as measure of the bias of the prediction from the reserving model, as
\small
$$
 E \left(\sum_{i=1}^{N^{R}(\tau)} \frac{Y_i-\hat{Y}_i}{\hat{\pi}_i(\tau)} \right) =  E \left(  \sum_{i=1}^{N(\tau)} \frac{Y_i-\hat{Y}_i}{\pi_i(\tau)} \mathbf{1}_i(\tau)\right) =  E \left(  \sum_{i=1}^{N(\tau)}Y_i-\hat{Y}_i \right) = - E \left(  \sum_{i=1}^{N(\tau)} \hat{Y}_i - \sum_{i=1}^{N(\tau)}Y_i \right) = -\textrm{Bias}\left(  \sum_{i=1}^{N(\tau)} \hat{Y}_i \right)
$$

\normalsize
and once it is added to construct the AIPW estimator, it subtracts the bias from the prediction of the model.  

Other approaches, which may not necessarily be superior, can also be used for this purpose. For example, imposing the balance property on the severity model can be viewed as a post-processing technique aimed at handling the bias in aggregate predictions. However, it is important to note that the way actuaries typically implement the balance property may not adequately address the sampling design, as we explain next.

Given the sampling bias, imposing the balance property solely on the reported data does not guarantee that aggregate quantities from the severity model are unbiased for the entire population of claims. It only ensures unbiasedness within the population of reported claims. To ensure the balance property holds at the overall population level, it is essential to account for the sampling mechanism that determines both reported and not reported claims, rather than focusing exclusively on reported claims.

In this context, we can extend the definition of the balance property of the severity model to different population levels: reported claims, not reported claims, or the entire portfolio, depending on which group is the target. Excluding exposure weights, these balance properties can be formulated as follows:
$$
\sum_{i=1}^{N^{R}(\tau)} \hat{Y}_i = \sum_{i=1}^{N^{R}(\tau)} Y_{i}, ~ ~~ ~ ~ ~ ~ ~ ~ \sum_{i=1}^{N^{R}(\tau)} \frac{1-\hat{\pi}_i(\tau)}{\hat{\pi}_i(\tau)}\hat{Y}_i = \sum_{i=1}^{N^{R}(\tau)} \frac{1-\hat{\pi}_i(\tau)}{\hat{\pi}_i(\tau)}Y_{i}, ~ ~ ~ ~ ~ ~ ~ ~ ~\sum_{i=1}^{N^{R}(\tau)} \frac{\hat{Y}_i}{\hat{\pi}_i(\tau)} = \sum_{i=1}^{N^{R}(\tau)} \frac{Y_{i} }{\hat{\pi}_i(\tau)}
$$

The first type of balance property aims to remove the sampling bias in the aggregated claims for the population of reported claims, the second for not reported claims, and the third for the entire population of claims. Each of these properties represents the standard balance property but applies a change of measure using odds ratio weights to account for the specific target populations. In reserving, where the primary focus is on predicting the population of not reported claims, the second balance property is most appropriate, as it directly targets the quantity of interest, which is the aggregate liability of not-reported claims. This approach is expected to lead to more accurate predictions for not-reported claims, and justifies the weighting scheme that shows up in the AIPW estimator in Equation (\ref{AIPW_Y3}).

A simple strategy to enforce this balance property involves adjusting the severity model using a calibration coefficient. Let $\hat{Y}_i^{wBP} $ be the adjusted prediction of the severity model obtained as:
$$
\hat{Y}_i^{wBP} = b \hat{Y}_i,
$$
where $b$ is determined to guarantee the weighted balance property. Simple algebra shows that this coefficient should be set to:

\begin{equation}
b = \frac{\sum_{i=1}^{N^{R}(\tau)}  \frac{1-\hat{\pi}_i(\tau)}{\hat{\pi}_i(\tau)}Y_i }{\sum_{i=1}^{N^{R}(\tau)}  \frac{1-\hat{\pi}_i(\tau)}{\hat{\pi}_i(\tau)}\hat{Y}_i}.
\label{bp}
\end{equation}

This adjustment aims to align the mean of the severity model with the mean of the distribution of not reported claims, thereby improving reserve prediction accuracy. However, the theoretical guarantees depend on the method used to impose the balance property. Alternatives, such as auto-calibration or constrained optimization, are available, but the most effective approach remains uncertain. Indeed, there is a gap on the literature on this regard as well as the general benefits of the balance property.

\subsection{In-processing: Fitting models for the population of not reported claims}
\label{wlike}
Another approach to address sampling bias in reported claims and its impact on severity model predictions involves modifying the algorithm for parameter estimation. Typically, parameter estimation for models is achieved by solving an estimating equation, such as \( \sum_{i=1}^{N(\tau)} l(Y_i, \theta) = 0 \), where the estimating equation could be the gradient of the log-likelihood. To account for sampling bias while estimating the model, we can modify the estimating equation to reflect the sampling design.

A common method in the sampling literature is to use weighted estimating equations, where the weights are the inverse of the inclusion probability \(\hat{\pi}_i(t)\), as discussed in studies such as \cite{wooldridge2007inverse}. However, in reserving, the focus shifts from estimating the overall population distribution \(f(Y)\) to predicting claims for not reported data, i.e., \(f(Y \mid U > \tau-T)\). To achieve this, we propose minimizing a weighted loss function that accounts for the sampling mechanism, with slight modifications to target the population of not reported claims. Based on the IPW principle, the following weighted equation is unbiased for the loss function of the not reported claims population (see Corollary \ref{unbiased}):
$$
\sum_{i=1}^{N^R} \frac{1-\hat{\pi}_i(\tau)}{\hat{\pi}_i(\tau)} l(Y_i, \theta)
$$
Here, the parameters \(\theta\) should be chosen to minimize this weighted loss function instead of the unweighted version, and so providing an estimate of \(f(Y \mid U > \tau-T)\), which is more suitable for the reserving context. This approach is grounded in the literature on missing data and survey designs with non-response. For foundational theory, see generalized estimating equations (e.g., Section 5.4 of \cite{shao2008mathematical}). Recent applications of similar procedures in insurance can be found in \cite{peiris2024integration}, while \cite{bucher2024combined} discusses related concepts in reserving.

This method provides predictions with clear interpretability, in contrast to post-processing techniques, where model predictions are adjusted in less transparent ways. Additionally, it adjusts all model parameters, not just the intercept, as in the balance property-based method discussed in the previous subsection. The degree to which the model parameters differ from those obtained using the regular unweighted loss function depends on the associations between non-reporting and the explanatory variables. As a result, the fitted model better represents the distribution of not reported claims, yielding more accurate predictions compared to models focused on reported claims or the entire population. We illustrate this advantage with real data in the next section.

Finally, while this paper does not explore alternative fitting approaches in detail, other methods exist (e.g., \cite{kim2021statistical}). For example, one could use standard estimating equations while imposing the balance property as a constraint. However, such methods tend to be computationally intensive, and their statistical properties are more complex to derive.

\subsection{Pre-Processing: Synthetic data sets of not reported claims}
\label{synthetic}

A general approach to handling partially observed data involves estimating the population of not reported claims using data augmentation, generation, or imputation methods. By augmenting the data from reported claims, we can create a pseudo-population of not reported claims that follows the distribution \(f(Y \mid U > \tau-T)\). This pseudo-population can then serve as the working data for subsequent analyses, including model fitting. This pre-processing method directly addresses the sampling bias in the data, rather than correcting it in later stages of estimation or prediction.

There are several methods to achieve this, and the following approach, commonly used for bootstrapping in the population sampling literature (e.g., \cite{arnab2017survey}), offers a straightforward solution. Briefly, each claim \(Y_i\) in the population appears, on average, \(\hat{\pi}_i\) times in the sample. Conversely, each \(Y_i\) in the sample is expected to appear \(1/\hat{\pi}_i\) times in the entire population of claims, or \(\frac{1-\hat{\pi}_i}{\hat{\pi}_i}\) times in the population of not reported claims.

Along those lines, synthetic data for not reported claims can be constructed by replicating each reported claim \((\mathbf{x}_i, Y_i)\) in a new dataset a total of \(\frac{1-\hat{\pi}_i}{\hat{\pi}_i}\) times. The synthetic data should closely mimic the not reported claims, allowing for accurate reserve estimation and related quantities. The following proposition shows this more formally.

\begin{proposition}
The distribution function of the severity of the resulting synthetic data of not reported claims is a strongly consistent estimation of the severity distribution of the population of not-reported claims.
\end{proposition}

\begin{proof}
The distribution function for the severity of this pseudo-dataset is immediately given as:
$$
 \hat{F}_Y(y) = \frac{ \sum_{i=1}^{N^R(\tau)}  \frac{1-\hat{\pi}_i(\tau)}{\hat{\pi}_i(\tau)} \mathbf{1}_{ \{Y_i \le y\}} }{ \sum_{i=1}^{N^R(\tau)}  \frac{1-\hat{\pi}_i(\tau)}{\hat{\pi}_i(\tau)} },
$$
which is a self-normalized IPW estimator ( or equivalently a self-normalized importance sampling estimator) of the CDF of the severity of not reported claims. Here the odds ratio acts as the appropriate weighting for the change of measure from the conditional distribution of reported claims to that of not reported claims, as discussed after Proposition \ref{IPWres}. This estimator is known to be a strongly consistent estimator (see e.g, Section 10.4 of \cite{ross2022simulation}) for the conditional distribution, which is  $f(Y \mid U > \tau-T)$. 
\end{proof}

The synthetic data can be used in various ways to aid the estimation of reserves. One approach is to fit the severity model using standard procedures without needing post- or pre-processing corrections, as the synthetic data already reflects the distribution of not reported claims. The resulting fitted model is closely related to one obtained via weighted estimating equations, as the loss function evaluated on the synthetic dataset mirrors the weighted loss function applied to the original data. Consequently, both approaches yield equivalent models. Alternatively, the synthetic data can be used to estimate the reserve directly by aggregating the claims and using the total as a point estimator. This approach aligns closely with the IPW estimator in Equation (\ref{IPW_Y1}) and produces identical predictions, as both methods use the same weighting scheme. However, working with synthetic data is often more practical, as it allows for the direct use of standard software implementations without modifications. Moreover, they enable analysis of claim size distributions, exploration of attribute associations, and incorporation of uncertainty into calculations as we explain next.

 While the process discussed earlier creates a single pseudo-population, multiple synthetic datasets can be generated to account for population uncertainty. For example, in the context of Poisson sampling in reserving, each reported claim can be viewed as a success in a Bernoulli trial, and therefore the number of attempts required to observe such a success follows a geometric distribution.

Along those lines, we propose to generate realization of random variables \(Z_i \sim \textrm{Geom}(\hat{\pi}_i), \quad i=1, \ldots, N^R(\tau)\),  to represent the number of not reported claims with the same characteristics as \(Y_i\). The synthetic dataset is then constructed by including each  \(Y_i\) a total of \(Z_i\) times. This approach allows for multiple simulations, where \(Z_i\)   acts as a random weight, analogous to the fixed weight \(\frac{1-\hat{\pi}_i}{\hat{\pi}_i}\). The pseudo-population created with fixed weights can be viewed as the "average" of these simulations. Although this geometric distribution-based method for synthetic data generation is uncommon, it has roots in missing data literature, such as the EM algorithm for truncated data (see Section 2.8 of \cite{mclachlan2007algorithm}). The proposition below validates the use of this simulation approach from a Bayesian perspective.

\begin{proposition}
Let $M_i(\tau)$ be the total number of claims with the same characteristics as $Y_i$. Let \( M_i^{R}(\tau) \) and \( M_i^{IBNR}(\tau) \) represent, respectively, the number of these claims that are reported before time \( \tau \) and the number that remain not reported by \( \tau \), which arise according to the Poisson sampling scheme with inclusion probability $\pi_i(\tau)$. If $M_i(\tau)$ follows an improper distribution as $P(M_i(\tau) = m) \propto 1/m$, then 
$$
M_i^{IBNR}(\tau) \mid M_i^{R}(\tau) \sim \textrm{NegBinom}(r=M_i^{R}(\tau), p=\pi_i(\tau) ).
$$
\end{proposition}
\begin{proof}
Because of the Poisson sampling scheme, we have  $M_i^{R}(\tau) \mid M_i(\tau) \sim \textrm{Binom}(n=M_i(\tau), p=\pi_i(\tau) )$. Therefore,
\begin{align*}
P( M_i^{IBNR}(\tau) = k \mid M_i^{R}(\tau) = r) & = P( M_i(\tau) = k + r \mid M_i^{R}(\tau) = r) \\
& \propto P( M_i^{R}(\tau) = r \mid M_i(\tau) = k + r ) \times P(M_i(\tau) = k + r) \\
& \propto \binom{k+r}{r} (\pi_i(\tau))^{r} (1-\pi_i(\tau))^{(k+r)-r} \times \frac{1}{k+r}\\
& \propto \binom{k+r-1}{k} (1-\pi_i(\tau))^{k} (\pi_i(\tau))^{r} .
\end{align*}
The proof finishes by noting that the last line is proportional to the pmf of the negative binomial distribution mentioned earlier.
\end{proof}

Note that while the proposition above refers to a negative binomial distribution, in practice, we use the geometric distribution because each reported claim is treated individually, as they all have entirely different characteristics. Moreover, the use of an improper prior distribution is motivated by Bayesian inference, where is known that improper prior acts as noninformative priors and thus minimizes its influence on the posterior, allowing the data to dominate and making the inference as data-driven as possible. Thus, from the perspective of the resulting posterior distribution, the improper prior is not a strong assumption. Additionally, other improper prior distributions can be used, such as a uniform, which also leads to a negative binomial distribution but with slightly different parameters. Nonetheless, the reciprocal form used in the proposition is widely used in Bayesian statistics and is particularly intuitive in the insurance context, as larger claims become increasingly less likely.

This method has several applications in reserving. It can quantify prediction uncertainty, similar to bootstrapping but focused on not reported claims, and generate confidence intervals for predictions and parameters without relying on asymptotic theory. For instance, since the average total claims from these datasets match the IPW estimator, bootstrap confidence intervals for the IPW can be constructed using quantiles from the synthetic totals generated in the simulations. Another application is model testing, as synthetic data provides a model-free method to evaluate prediction quality. Furthermore, for micro-level models that require known claim features (often unavailable), this approach offers a straightforward way to generate claims and their attributes.

\section{Illustration with real data}
\label{numerical}
In this section, we examine the performance of the various sampling-based approaches described earlier and compare them with traditional methods using a real dataset from a Canadian automobile insurance company. Because this dataset is proprietary and cannot be shared, we have also prepared a small simulation study—including the corresponding code for implementation—to illustrate the methodology. This is available at \url{https://htmlpreview.github.io/?https://github.com/sfcalcetero/AIPW/blob/main/CIA-report--F.html}. This simulation is intended for practical demonstration rather than serving as a comprehensive case study.

The Canadian dataset for this section contains detailed records of Physical Damage (PD) claims from January 2014 to December 2016, documenting individual claims, policyholder demographics, and automobile specifications. The information includes vehicular attributes such as age, horsepower, weight, and market value; policyholder characteristics like gender and age; and accident-specific details, including the time of occurrence, time of reporting, and the claim amount.

\begin{figure}[h]
        \centering
        \includegraphics[width=0.6\textwidth]{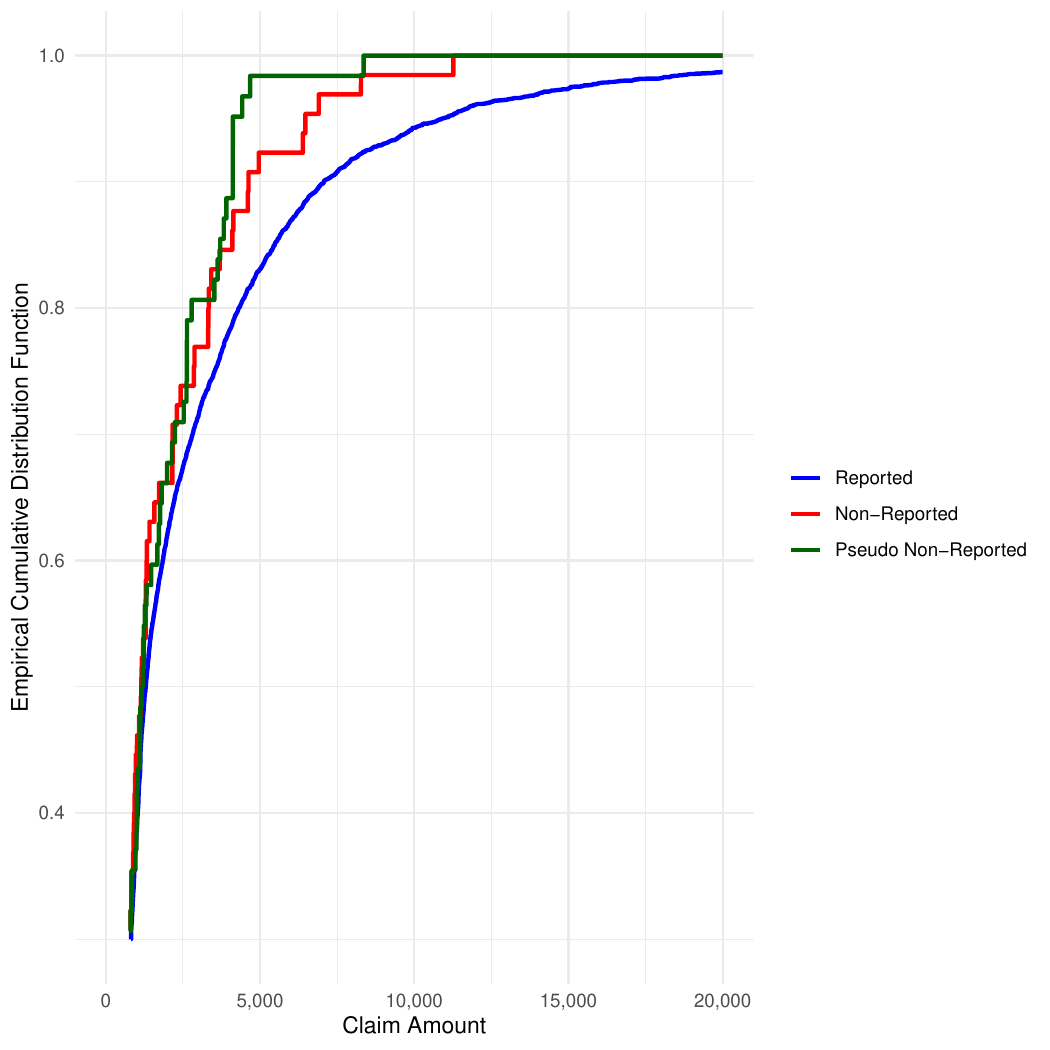}         
     \caption{Empirical CDFs of the claim severity of reported and not reported claims for the first month in the testing set}
     \label{ecdf}
\end{figure}

A key characteristic of this dataset is the significant difference in the claim amounts between reported and not reported claims, with not reported claims generally having lower severity. This discrepancy is common in auto insurance and is largely due to reporting behaviors. Larger, more severe claims are often reported immediately, leading to an over-representation of high-severity claims among those reported. In contrast, smaller claims, which may involve minor damages, are more likely to be delayed or not reported, resulting in a lower average severity for not reported claims. Figure \ref{ecdf} illustrates the distributions of the claim amounts for both groups, showing that reported claims tend to have a longer tail than not reported ones. 

Along those lines, loss estimation methods based solely on reported claims tend to overestimate reserves if this difference is not accounted for. To preview how the population sampling approach can address this issue, we also plot in Figure \ref{ecdf} the distribution of a pseudo-population of not reported claims, synthetically generated using the methodology described in Section \ref{synthetic}. This synthetic data set more closely resembles the distribution of true not reported claims than what would be obtained from reported claims alone.

In the following, we will estimate the reserves of not reported claims on a monthly basis, end of the month, using the period of time 2014 and 2015 as training and 2016 as the testing period.

\subsection{Model fitting}
\label{fitting}
To effectively illustrate the improvements gained from the sampling methods, we need a model to assist in the estimation. For this purpose, we first fit a micro-level model to our data. The model described below yielded the best goodness-of-fit among various distributional assumptions. For practical application, we refit the model every month to incorporate the latest available data, which means the model parameters change over time. The parameter values presented here are for the first month of the testing period, serving as an illustration.

\subsubsection*{Reporting delay time and inclusion probabilities}
Here we estimate a single model that would be used for both the reporting delay time and the construction of the inclusion probabilities. As mentioned earlier, inclusion probabilities are a critical component for the application of the population sampling methods, so special attention is given to this step. 
 
Along those lines, we use a flexible model that accounts for the right-truncation of the reporting delay time. The structure selected is the one of a Cox regression model that describes the hazard function of the reporting delay time, $\lambda_{U \mid \boldsymbol{x} }(u)$, using the attributes of the policyholder $\boldsymbol{x}$ as covariates, as follows:
\small
\begin{align*}
\log( \lambda_{U} (u)  ) =&  \log(\lambda_0(u)) +\beta_1\mathrm{Car\_Age}+\beta_2\mathrm{Claim\_Count}+\beta_3\mathrm{log(Horse\_Power)}+\beta_4\mathrm{log(Car\_Weight)} \\
&+\beta_5\mathrm{log(Car\_Price)}+\beta_6\mathrm{Gender}+\beta_7\mathrm{Driver\_Age}+S_8(\mathrm{Accident\_day})+S_9(\mathrm{Claim\_Amount}).
\end{align*}
\normalsize

The log of the baseline hazard function, $\log(\lambda_0(u))$, is estimated using a B-Spline representation. Additionally, we incorporate nonlinear effects for the covariates "Accident\_day" and "Claim\_Amount" with the terms $S_8(\textrm{Accident-day})$ and $S_9(\mathrm{Claim\_Amount})$ respectively, which are also estimated using a B-Spline representation.  The implementation is done via a generalized additive model with a piecewise exponential function that accounts for right-truncation. This is available in \texttt{R} packages such as \texttt{flexsurvreg}, \texttt{pammtools} or \texttt{reservr}. 

\begin{figure}[h]
     \centering
     \begin{subfigure}[b]{0.32\textwidth}
         \centering
         \includegraphics[width=\textwidth]{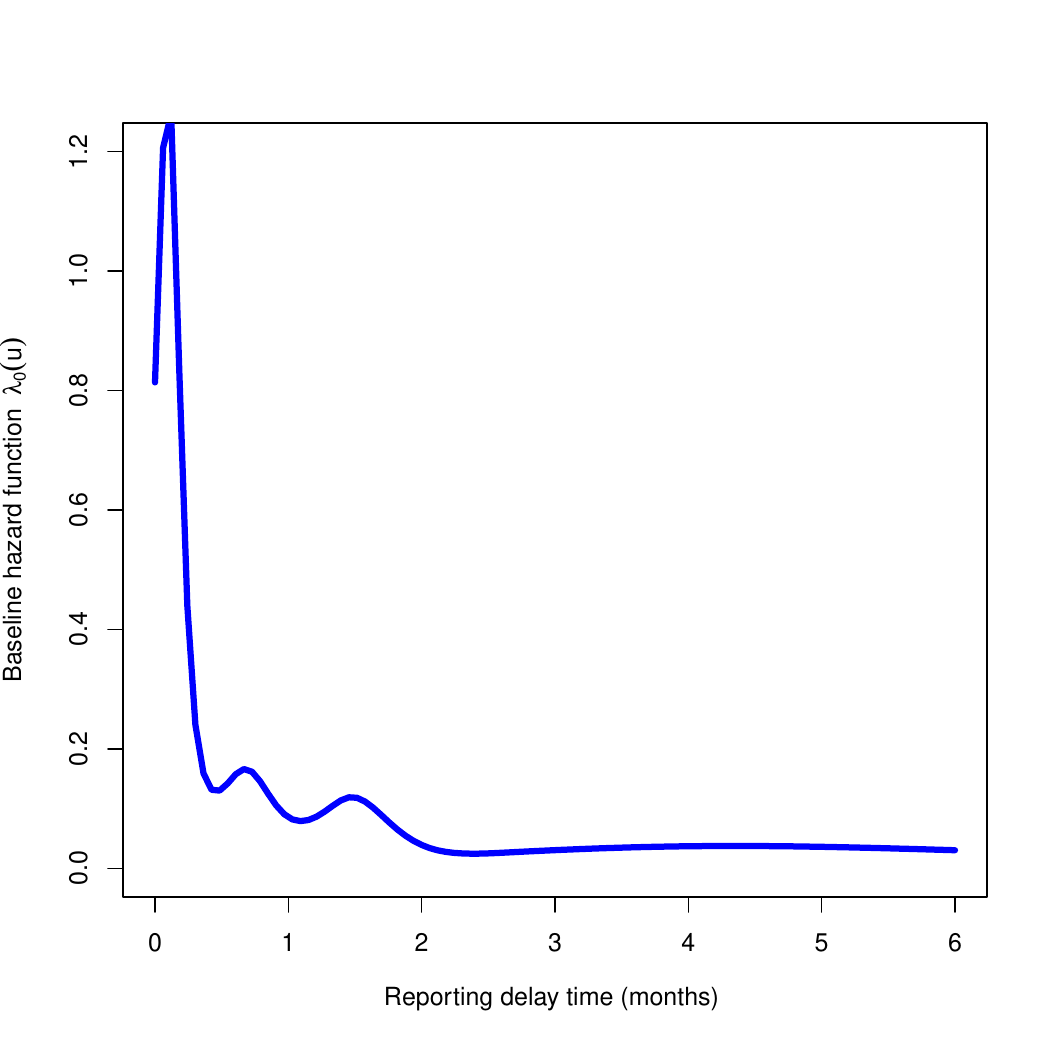}
        \end{subfigure}
     \hfill
     \begin{subfigure}[b]{0.32\textwidth}
         \centering
        \includegraphics[width=\textwidth]{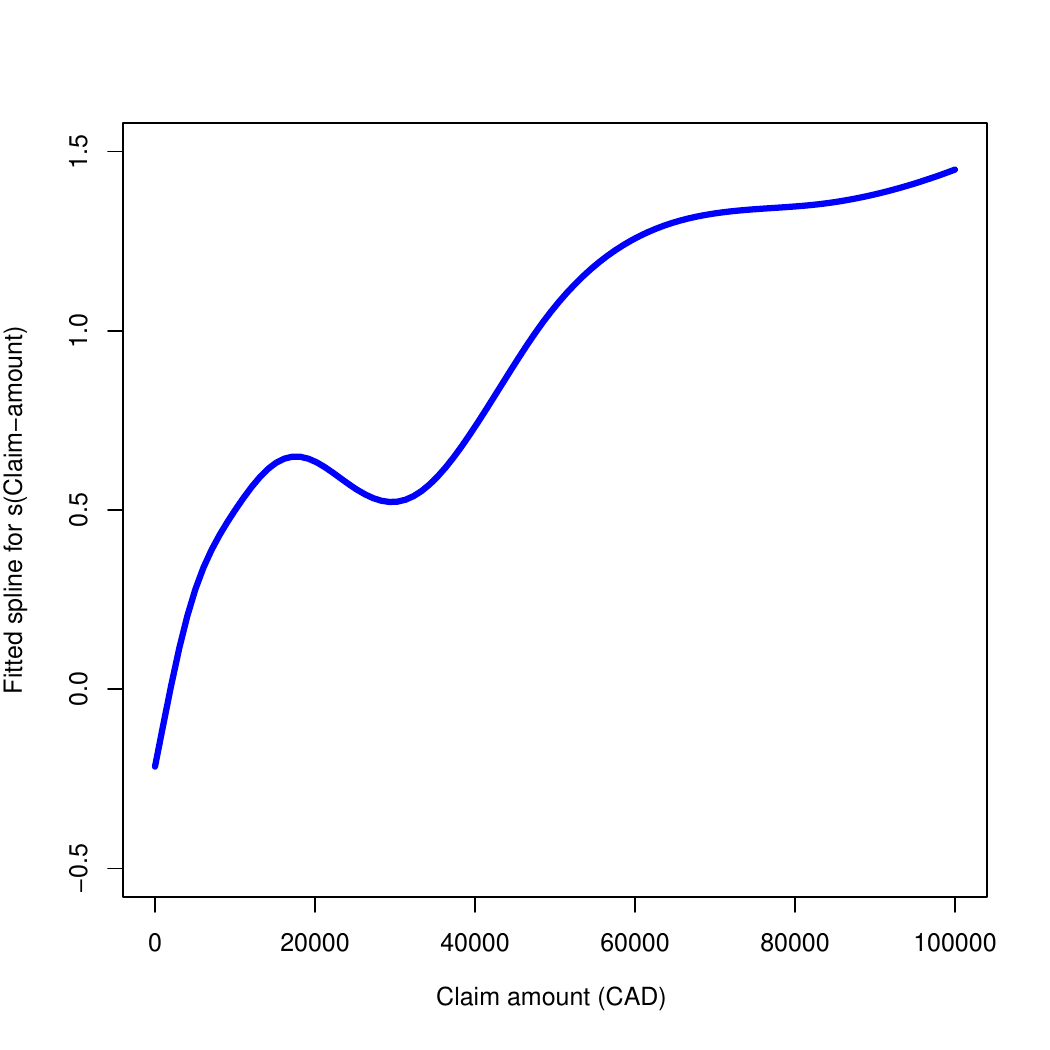}          
     \end{subfigure}
     \hfill
     \begin{subfigure}[b]{0.32\textwidth}
         \centering
        \includegraphics[width=\textwidth]{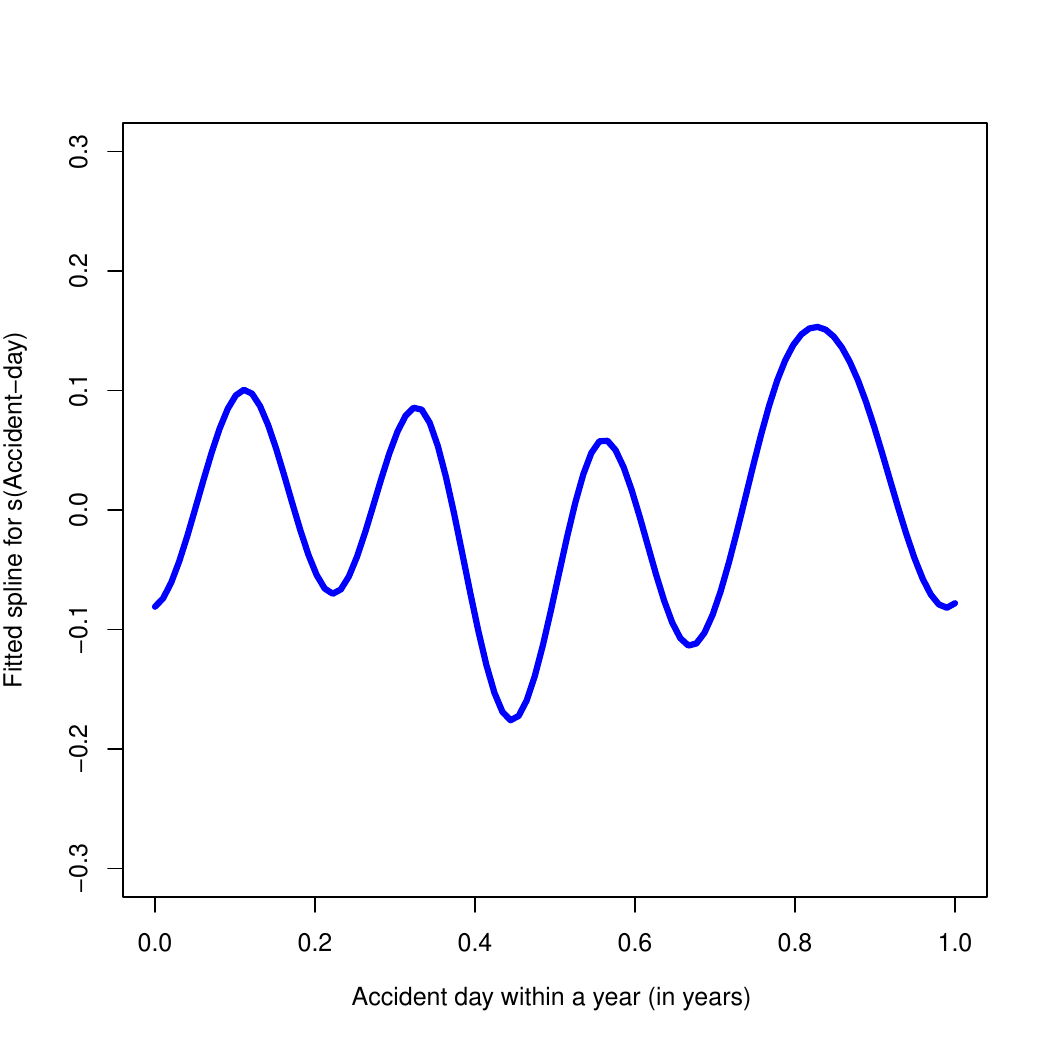}          
     \end{subfigure}
     \caption{ Fitted baseline hazard function for the reporting delay time  (left panel)  and fitted  effects of the  claim amount (center panel) and the accident day on the hazard of the reporting delay time (right panel) }
     \label{fitted_repo}
\end{figure}

The fitted parameters are shown in Table \ref{regcoef_delay} and Figure \ref{fitted_repo}. All regression coefficients are statistically significant. The baseline hazard function indicates that most claims are reported within the first two months, but there is a long tail for claims with extended reporting times. Furthermore, the claim size has an almost increasing effect, suggesting that larger claims tend to be reported more quickly. This is consistent with the observed greater severity for reported claims than for not reported. Finally, the effect of accident day on the hazard function reveals a quarterly seasonality.

\begin{table}[h]
\small
\centering
\resizebox{0.9\textwidth}{!}{
\begin{tabular}{cccccccccc}
\hline
\hline 
\textbf{Coefficient}         & \textbf{$\beta_1$} & \textbf{$\beta_2$} & \textbf{$\beta_3$} & \textbf{$\beta_4$} & \textbf{$\beta_5$} & \textbf{$\beta_6$} & \textbf{$\beta_7$}   \\ \hline
Value    & $7.08 $  & $-17.41 $ & $4.28 $  & $-38.62 $ & $-255.88 $ & $36.46 $  & $-2.03 $ \\
Std. Err. & $0.07 $  & $0.38 $  & $1.66 $  & $2.09 $  & $1.06 $  & $0.49 $  & $0.01 $ \\
p-val  & $<0.001$ & $<0.001$ & $<0.001$ & $<0.001$           & $<0.001$  & $<0.001$ & $<0.001$   \\ \hline
\hline
\end{tabular}
}
\caption{Estimated regression coefficients of the Cox regression model for the reporting delay time, scaled by $\times 10^{3}$}
\label{regcoef_delay}
\end{table}

Lastly, we use the model to obtain the inclusion probabilities, which can be computed by plugging in the estimated hazard function into the expression below:
\begin{equation}
\hat{\pi}_i(\tau) = Pr(U_i \le \tau-T_i \mid \boldsymbol{x}_i, T_i, Y_i)= 1-\exp( -\int_0^{\tau-T_i} \lambda_{U }(u)  du).
\end{equation}

\FloatBarrier
\subsubsection*{Frequency model}
We found that the usual  Poisson process assumption for the number of claims per policyholder did not fit well due to significant overdispersion and an excess of zero claims. To address this, we used a practical approach by fitting a model for the random variable representing the number of claims up to the valuation time \(\tau\) and applied thinning to derive the distribution of not reported claims.

In our dataset, the Zero-Inflated Negative Binomial (ZINB) model provided a good fit. The ZINB has the following probability mass function
$$
P_{ZINB}(k; q, \theta, r) := q\mathbf{1}_{ \{ k = 0 \} }+(1-q)\binom{k + r - 1}{k} \left(\frac{\theta}{\theta + r}\right)^k \left(\frac{1}{\theta + r}\right)^r, 
$$
\noindent where \(q\) is the probability of zero inflation, \(\theta\) is associated with the mean of the negative binomial distribution, and \(r\) is a dispersion parameter. Along those lines, the frequency model is the following: 

\begin{itemize}
\item Let the total number of claims from the $j$-th contract up to \(\tau\), \(N_j(\tau)\), follow a ZINB regression:
$$
P(N_j(\tau) =n) = P_{ZINB}(n; q_j, \xi_j\theta_j, r),
$$ \begin{align*}
logit(q_j) = &\beta_0+\beta_1\mathrm{Car\_Age}+\beta_2\mathrm{Claim\_Count}+\beta_3\mathrm{log(Horse\_Power)}+\beta_4\mathrm{log(Car\_Weight)}\\
&+\beta_5\mathrm{log(Car\_Price)}+\beta_6\mathrm{Gender}+\beta_7\mathrm{Driver\_Age}+\beta_8 \xi, \\
\log(\theta_j) = &\beta_0+\beta_1\mathrm{Car\_Age}+\beta_2\mathrm{Claim\_Count}+\beta_3\mathrm{log(Horse\_Power)}+\beta_4\mathrm{log(Car\_Weight)}\\
&+\beta_5\mathrm{log(Car\_Price)}+\beta_6\mathrm{Gender}+\beta_7\mathrm{Driver\_Age}
\end{align*}
where \(r\) is a common dispersion parameter, and \(\xi_j\) is a known exposure term.

\item The arrival time \( T \) of a claim has cumulative distribution function \( F_T(t) \), which may depend on covariates. In our dataset, and commonly observed in reserving, accident times are uniformly distributed over the life of the contract, so we use this distribution.

\item Assuming independence of delays for each claim, the marginal distributions for reported claims and not reported claims are given by:
\begin{align}
\label{Dist_R} 
P(N_j^R(\tau)=k) = P_{ZINB}(k; q_j, p_j(\tau) \xi_j \theta_j, r) \\ 
P(N_j^{IBNR}(\tau)=k) = P_{ZINB}(k; q_j, (1-p_j(\tau)) \xi_j \theta_j, r) 
\end{align}
\noindent where \(p_j(\tau)=\int_0^{\tau} \hat{\pi}_i(\tau-t)dF_T(t)\) is the average inclusion probability. For uniform accident times, this simplifies to \(p_j(\tau)=\frac{1}{\tau} \int_0^{\tau} \pi_j(\tau-t)dt\).

\item For prediction, we can use the conditional distribution of not reported claims given reported claims. By Bayes' rule, we find that 
\begin{equation*}
P(N_j^{IBNR}(\tau)=k \mid N_j^R(\tau)) = P_{ZINB}(k; \tilde{q}_j, (1-p_j(\tau)) \xi_j \tilde{\theta}_j, \tilde{r}_j) 
\end{equation*}

where \(\tilde{q}_j= \frac{q_j \mathbf{1}_{ \{N^R_j(\tau)=0 \} }}{q_j+(1-q_j)(1+\theta_j\xi_jp_j(\tau))^{-r}}\), \(\tilde{\theta}_j = \frac{\theta_j}{1+\theta_j\xi_jp_j(\tau)}\), and \(\tilde{r}_j = r+N^R_j(\tau)\). Thus, the expected number of not reported claims for the \(j\)-th policyholder, denoted as \(\lambda_j^{IBNR}\), is:
\begin{equation}
\lambda_j^{IBNR} = (1-\tilde{q}_j) \tilde{r}_j (1-p_j(\tau)) \xi_j \tilde{\theta_j} 
\label{Preditive_IBNR}
\end{equation}

This approach has not been used in reserving before and offers a flexible alternative to the Poisson process while remaining analytically tractable. Unlike the Poisson process, this method incorporates the current number of reported claims into the prediction, similar to the dynamic claim score by \cite{yanez2024modeling}, which considers historical claiming behavior.

\end{itemize}

Using the reported claims, we fit the ZINB regression for reported claims as in Equation (\ref{Dist_R}) via maximum likelihood, and then plug the estimated parameters in Equation (\ref{Preditive_IBNR}) for predictions. This can be  done with \texttt{R} packages for GLMs, such as \texttt{pscl}, where the term $\xi_jp_j(\tau)$ acts as an offset. The fitted model for the first month in the testing period is:

\begin{table}[H]
\small
\centering
\resizebox{\textwidth}{!}{
\begin{tabular}{ccccccccccc}
\hline
\hline 
\textbf{} & \textbf{Coefficient}   & \textbf{$\beta_0$}       & \textbf{$\beta_1$} & \textbf{$\beta_2$} & \textbf{$\beta_3$} & \textbf{$\beta_4$} & \textbf{$\beta_5$}  & \textbf{$\beta_6$}  & \textbf{$\beta_7$} & \textbf{$\beta_8 \textrm{ \& } r$} \\ \hline
& Value   & $1.13$  & $0.10$  & $-0.25$ & $0.28$  & $0.74$ & $-1.09$ & $0.19$  & $0.01$ & $1.31$ \\
$logit(q)$ & Std. Err. & $2.68$  & $0.02$  & $0.09$  & $0.43$  & $0.61$  & $0.35$  & $0.12$  & $0.00$  & $0.06$ \\
& p-val & $<0.001$ & $<0.001$ & $<0.001$ & $<0.001$ & $<0.001$           & $<0.001$  & $<0.001$ & $<0.001$ & $<0.001$  \\ \hline
& Value  & $-1.82$  & $-0.02$  & $0.05$ & $0.13$  & $0.22$ & $-0.42$ & $0.12$  & $0.00$ & $16.37$ \\
$\log(\theta)$ & Std. Err. & $1.61$  & $0.01$  & $0.05$  & $0.25$  & $0.34$  & $0.19$  & $0.07$  & $0.00$ & - \\
& p-val & $<0.001$  & $<0.001$ & $<0.001$ & $<0.001$ & $<0.001$  & $<0.001$  & $<0.001$  & $<0.001$ & - \\ \hline

\hline
\end{tabular}
}
\caption{Estimated regression coefficients of the ZINB model for the number of claims}
\label{regcoef_freq}
\end{table}

\subsubsection*{Severity models}
 
This is the main component of the reserving model where the sampling methods play a crucial role. As noted earlier, the severity distribution differs significantly between reported and not reported claims. Therefore, we fit three versions of the severity model under different configurations: one using the regular approach (i.e., without incorporating the sampling design into the estimation), and the other two using population sampling adjustments.

For the general regression model, we found that a lognormal regression  was well-suited to our data, as specified below. We will also assume that claim amounts are iid for a given contract.
\begin{align*}
&  \log(Y_i) \sim Norm (\nu_i, \sigma^2) \\
\nu_i = &\beta_0+\beta_1\mathrm{Car\_Age}+\beta_2\mathrm{Claim\_Count}+\beta_3\mathrm{log(Horse\_Power)}+\beta_4\mathrm{log(Car\_Weight)}\\
&+\beta_5\mathrm{log(Car\_Price)}+\beta_6\mathrm{Gender}+\beta_7\mathrm{Driver\_Age}
\end{align*}
This model can be easily fitted using a least-squares procedure, such as \texttt{lm} in \texttt{R}, on the log-scale of the claim amounts. The  prediction, given by the mean, is then:
$$
\hat{Y}_i = \exp(\nu_i+\sigma^2/2).
$$

Now, we proceed to estimate the model and the respective adjustments based on sampling theory. Conceptually, we are working with the same model specification, i.e., the same distribution and regression formula as mentioned above. However, these variations will differ in the estimated parameters as follows:

\begin{itemize}

\item The first estimation assumes that severity and reporting are independent. In this case, we use maximum likelihood as usual, which coincides with minimization using unweighted least squares. We denote the point prediction as \(\hat{Y}_i\). The fitted parameters are:

\begin{table}[H]
\small
\centering
\resizebox{0.9\textwidth}{!}{
\begin{tabular}{cccccccccc}
\hline
\hline 
\textbf{Coefficient}   &  \textbf{$\beta_0$}    & \textbf{$\beta_1$} & \textbf{$\beta_2$} & \textbf{$\beta_3$} & \textbf{$\beta_4$} & \textbf{$\beta_5$} & \textbf{$\beta_6$} & \textbf{$\beta_7$} & \textbf{$\sigma$}  \\ \hline
Value    & $7.10$  & $-0.03$  & $-0.04$ & $0.01$  & $-0.44$ & $0.36$ & $0.03$  & $-0.002$ & $0.73$ \\
Std. Err. & $0.68$  & $0.00$  & $0.02$  & $0.10$  & $0.14$  & $0.07$  & $0.03$  & $0.0001$ \\
p-val  & $<0.001$ & $<0.001$ & $<0.001$ & $<0.001$ & $<0.001$ & $<0.001$          & $<0.001$  & $<0.001$   \\ \hline
\hline
\end{tabular}
}
\caption{Estimated regression coefficients of the log-normal model for the claim amount}
\label{regcoef_severity}
\end{table}

\item The second model is estimated using the weighted loss function approach, as proposed in Section \ref{wlike}. This no longer imposes independence with reporting on the estimated model. In this case, instead of maximizing directly the log-likelihood, we maximize its weighted version, i.e. $\sum_{i=1}^{N^R} \frac{1-\hat{\pi}_i(\tau)}{\hat{\pi}_i(\tau)} \log f(Y_i)$, which coincides with a weighted least squares procedure for our selected model.  The fitted parameters are as follows:

\begin{table}[H]
\small
\centering
\resizebox{0.9\textwidth}{!}{
\begin{tabular}{cccccccccc}
\hline
\hline 
\textbf{Coefficient}   &  \textbf{$\beta_0$}    & \textbf{$\beta_1$} & \textbf{$\beta_2$} & \textbf{$\beta_3$} & \textbf{$\beta_4$} & \textbf{$\beta_5$} & \textbf{$\beta_6$} & \textbf{$\beta_7$} & \textbf{$\sigma$}  \\ \hline
Value    & $7.98$  & $-0.01$  & $-0.01$ & $0.31$  & $-0.99$ & $0.47$ & $0.03$  & $-0.0001$ & $0.84$ \\
Std. Err. & $0.48$  & $0.003$  & $0.02$  & $0.07$  & $0.09$  & $0.05$  & $0.05$  & $0.001$ \\
p-val  & $<0.001$ & $<0.001$ & $<0.001$ & $<0.001$ & $<0.001$ & $<0.001$          & $<0.001$  & $<0.001$   \\ \hline
\hline
\end{tabular}
}

\caption{Estimated regression coefficients of the log-normal model for the claim amount using weighted least squares}
\label{regcoef_severity_weight}
\end{table}

Although the estimated parameters show similarities with the first model, the values differ. This highlights the differences between the severity of reported and not reported claims. Recall that this estimated model is equivalent to the one that would be estimated if we used the synthetic data set. As such, we do not make a distinction and will denote the point prediction as \(\hat{Y}_i^{WL}\). Nevertheless, the synthetic data set approach may be used to construct bootstrapped versions of the coefficients.

\item The third model is a variation of the first, where we impose the weighted balance property as described in Section \ref{dbp}. We denote the point prediction as \(\hat{Y}_i^{wBP}\). The fitted parameters are the same as those of the first model, except the intercept of the models, which changes via the multiplicative coefficient from Equation (\ref{bp}) as:
$$
\hat{Y}_i^{wBP} = \left( \frac{\sum_{i=1}^{N^{R}(\tau)}  \frac{1-\hat{\pi}_i(\tau)}{\hat{\pi}_i(\tau)}Y_i }{\sum_{i=1}^{N^{R}(\tau)}  \frac{1-\hat{\pi}_i(\tau)}{\hat{\pi}_i(\tau)}\hat{Y}_i} \right)\hat{Y}_i =  0.70\hat{Y}_i
$$
As the coefficient is less than 1, the adjusted model will predict a reduced severity when compared to the model that assumes independence. Note that this is consistent with the fact that the severity of not reported claims tends to be lower than that of reported claims.

\end{itemize}

\subsection{Estimation of the reserve }

\begin{figure}[h]
        \centering
        \includegraphics[width=\textwidth]{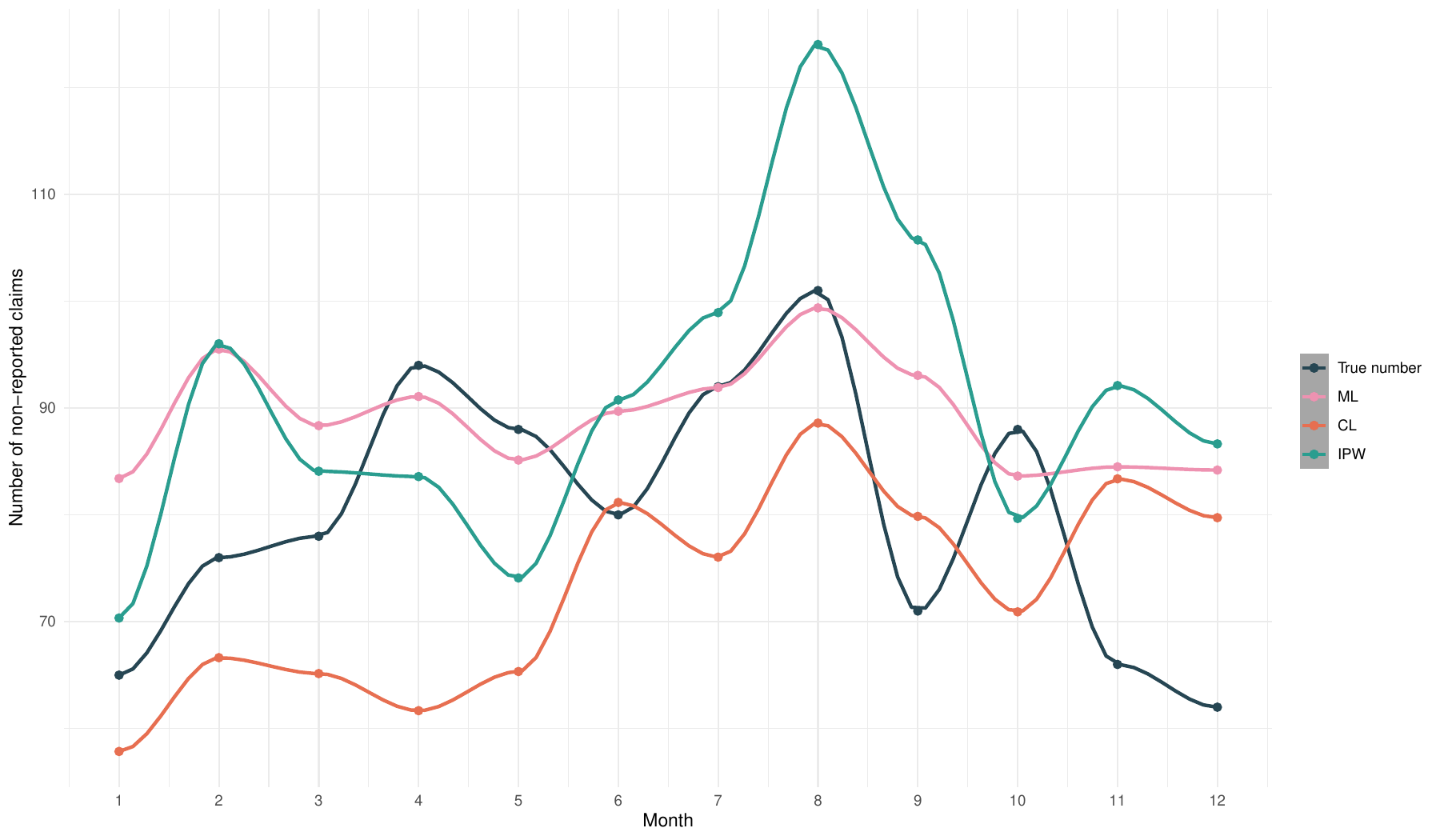}         
     \caption{Estimation for the number of not reported claims per month }
     \label{plot_tot_n}
\end{figure}

\begin{figure}[h]
        \centering
        \includegraphics[width=\textwidth]{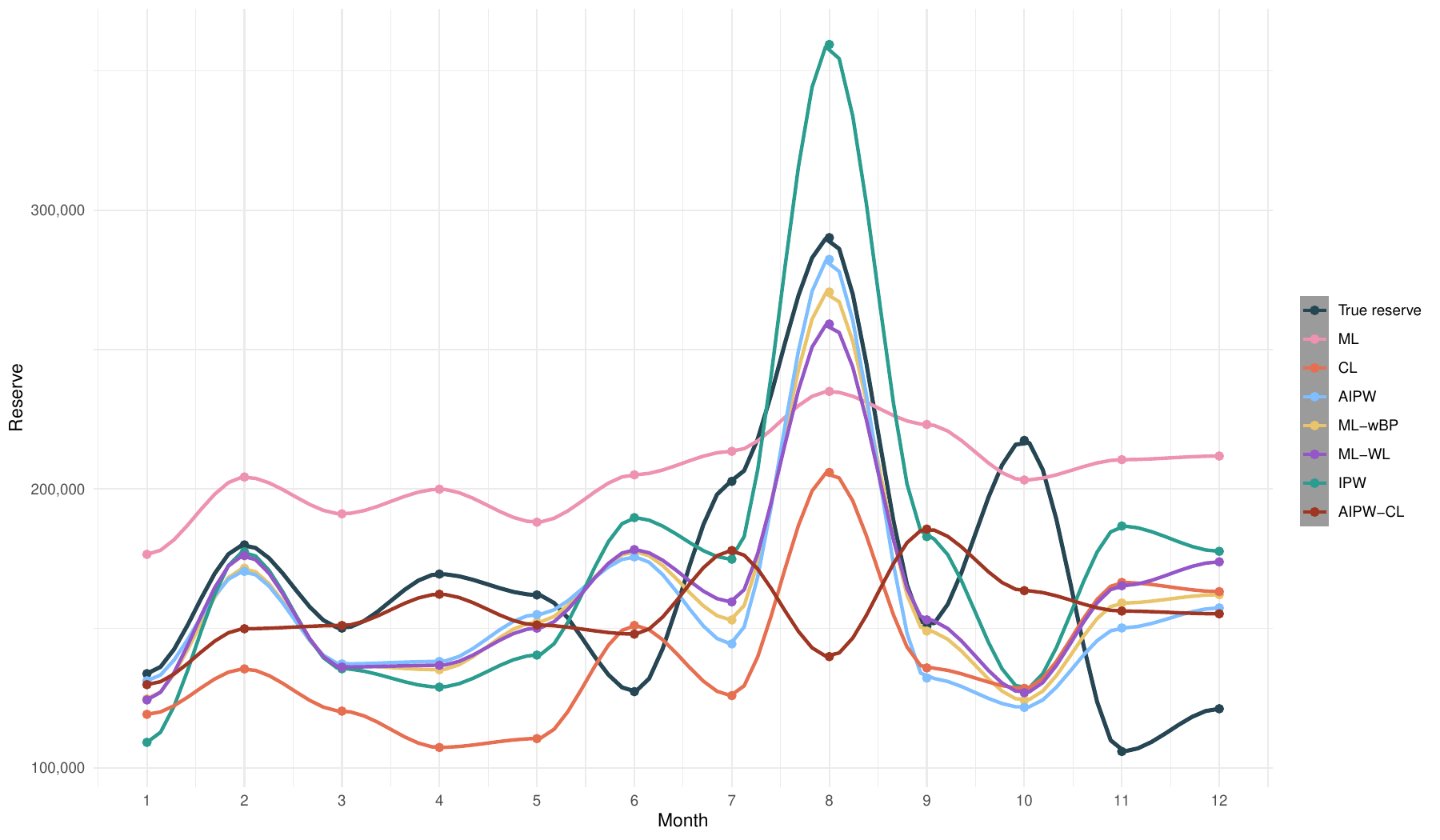}         
     \caption{Estimation for the outstanding claims reserve per month }
     \label{plot_tot}
\end{figure}

Here we present the reserve estimations during the testing period. We compare the actual reserve values, both in terms of amount and number, with the predictions from various models detailed in Table \ref{table:models}. These models incorporate different combinations of components from the previous section, and the results are shown in Figures \ref{plot_tot} and \ref{plot_tot_n}, as well as Table \ref{tab_err}. 

The models selected for comparison in Table \ref{table:models} are categorized based on the nature of their estimation methods. The primary objective is to demonstrate how the sampling-motivated reserve estimators can outperform traditional methods, highlighting any potential improvements. To this end, the first two models are classic reserving methods, serving as a baseline for comparison. The third, fourth and fifth models showcase the main sampling-motivated modifications for addressing bias, including the AIPW estimator.  The last two models explore alternative structures of the AIPW that represent an intermediate approach between micro- and macro-models.  The predictions based solely on the synthetic dataset of not reported claims yield point estimations equivalent to the regular IPW estimator and are therefore not listed separately in the analysis.

\begin{table}[h]
\centering
\resizebox{0.95\textwidth}{!}{
\begin{tabular}{>{\centering\arraybackslash}m{3cm}>{\centering\arraybackslash}m{0.4\textwidth}>{\centering\arraybackslash}m{0.45\textwidth}}
\hline
\hline
\textbf{Model} & \textbf{Description} & \textbf{Expression for Prediction} \\
\hline
\hline
\multicolumn{3}{l}{ \footnotesize \textit{Traditional reserving models} }\\
ML & Prediction using the components of the micro-level model. & $\sum_{j=1}^M \lambda_j^{IBNR} \hat{Y}_j$ \\
\hline
CL & Prediction using the traditional development factors ($f_k(\tau) $) approach. & $\sum_{k=1}^m  \left( \sum_{i \mid T_i=k} Y_i \right) \times  \left(  f_k(\tau) -1 \right)$ \\
\hline
\hline
\multicolumn{3}{l}{  \footnotesize \textit{AIPW estimators}} \\
AIPW & Prediction of ML along with the augmentation term & $\sum_{j=1}^M \lambda_j^{IBNR} \hat{Y}_j+\sum_{i=1}^{N^{R}(\tau)} \frac{1-\hat{\pi}_i(\tau)}{\hat{\pi}_i(\tau)}(Y_{i}-\hat{Y}_i)$ \\
\hline
IPW &  Prediction using Equation (\ref{IPW_Y1}), with probs from the Cox model. a.k.a Micro-level CL. & $\sum_{i=1}^{N^{R}(\tau)} \frac{1-\hat{\pi}_i(\tau)}{\hat{\pi}_i(\tau)} Y_{i}$ \\
\hline
AIPW-CL & Same as AIPW, but using probabilities implied by the CL  & $\sum_{j=1}^M \lambda_j^{IBNR} \hat{Y}_j+\sum_{i=1}^{N^{R}(\tau)} \frac{1-\pi^{CL}_{i}(\tau)}{\pi^{CL}_{i}(\tau)}(Y_{i}-\hat{Y}_i)$ \\
\hline
\hline
\multicolumn{3}{l}{  \footnotesize \textit{Other sampling methods}} \\

ML-wBP & Same as ML, but with modified intercept for balance property as in Equation (\ref{bp}) & $\sum_{j=1}^M \lambda_j^{IBNR} \hat{Y}_j^{wBP}$ \\
\hline
ML-WL & Same as ML, but severity estimated with weighted likelihood & $\sum_{j=1}^M \lambda_j^{IBNR} \hat{Y}_j^{WL}$ \\

\hline
\hline
\end{tabular}
}
\caption{Models in the numerical comparison:  Traditional Reserving Models, AIPW estimators, and other methods. Note that ML-wBP can be thought of as an AIPW estimator as the associated augmentation term would be 0.}
\label{table:models}
\end{table}

Figure \ref{plot_tot_n} shows the point estimates for the number of not reported claims compared to the actual values for each month. These are obtained by setting $Y_i=\hat{Y}_i=1$ in all the formulas. As some of the methods listed in Table \ref{table:models} differ only in their severity component, they are not listed as they lead to identical predictions for the number of claims. The results suggest that the various methods are competitive in predicting the number of not reported claims during the testing period, with no consistent pattern of over- or underestimation. However, the CL method consistently underestimates the number of not reported claims. Overall, the methods generally capture the trend in fluctuations of not reported claims, with the ML model exhibiting behavior closest to the actual data. These findings imply that the fitted models for frequency and inclusion probabilities are appropriately specified, effectively capturing the nuances of claim counts and reporting delays. 

Figure \ref{plot_tot} presents the point estimates for the reserve generated by all methods compared to the true values for each month. The results indicate that the ML reserve consistently overestimates the reserve throughout the testing period. Given that the models for frequency and reporting delay appropriately fit the number of claims, this overestimation is therefore attributed to the severity component of the model. This finding aligns with the earlier observation that the mean severity of reported claims (which are used to fit the model) is higher than that of not reported claims.

\begin{table}[h]
 
\centering
\begin{tabular}{cccccccc}
\hline
\hline
 \textbf{Metric} & \textbf{ML} & \textbf{CL}  & \textbf{AIPW} & \textbf{ML-wBP} & \textbf{ML-WL} & \textbf{IPW} & \textbf{AIPW-CL}  \\ \hline
ME & -37,605 & 28,400  & 9,563 & 7,874 & \textbf{5,924} & -6,691 & 11,668 \\
MAE & 49.155 & 49,454  &  \textbf{31,016} & 31,921 & 33,476 & 43,482 & 35,131 \\
RMSE & 57,356 & 55,288  & \textbf{40,669} & 40,974 & 42,240 & 50,896 & 52,027\\
MAPE & 34.9\% & 29.3 \% & \textbf{19.6\%} & 20.4\% & 21.6\% & 28.2\% & \textbf{19.6\%}\\
\hline
\hline
\end{tabular}
\caption{Error metrics for the total of the reserves over the testing period. ME: Mean error, RMSE: Root mean square error, MAE: Mean absolute error, MAPE: Mean absolute percentage error. Bolded text indicates the model with the best performance on each row.}
\label{tab_err}
\end{table}

The performance of the ML model adjustments using the AIPW, ML-wBP, and ML-WL methods is notable. Figure \ref{plot_tot} and Table \ref{tab_err} show that these approaches yield nearly identical estimates, with slight differences in some months. This similarity, expected given their shared goal of bias correction, highlights the significant bias in the original ML model. The unbiased estimates are closer to the predictions of these methods, which effectively detect and adjust the ML model's positive bias. In most cases, the correction is downward, except in month 8, where an upward adjustment accounts for a peak in the reserve.

A key feature of these adjustments is their ability to enhance not only the point prediction but also the reserve's variability. While the ML model provides stable but overly smooth estimates over time, the adjusted methods capture the reserve's fluctuations more accurately. This ability to reflect the reserve's dynamic behavior underscores the value of integrating sampling theory-inspired bias correction techniques with micro-level models. Table \ref{tab_err} quantitatively demonstrates the improvement, with the AIPW estimator achieving the most significant error reduction.

Turning to the CL, IPW, and AIPW-CL methods, the CL approach underestimates the reserve but captures its overall trend, which is valuable from a macro-level perspective. The IPW estimator provides a closer approximation to the actual reserve while maintaining a similar trend. These two methods behave similarly because the CL approach simplifies the IPW estimator. By incorporating claim and policyholder attributes into the inclusion probabilities and development factors, the IPW method significantly improves the CL estimates. Notably, these methods do not rely on frequency/severity models, yet their reserve predictions remain competitive.

The AIPW-CL method combines elements of the ML model's stability with the CL method's ability to capture fluctuations. Its estimates, blending the strengths of both models, deliver more desirable predictions than either approach individually. This illustrates how the AIPW estimator integrates the corrective strengths of both methods into a unified prediction. As discussed in Section \ref{gap}, the IPW and AIPW-CL methods act as intermediaries between micro-level and macro-level models, leveraging the advantages of both frameworks. The AIPW estimator, striking a balance between the trend behavior of macro-level approaches and the accuracy of micro-level predictions, offers enhanced predictive power overall.

\section{Conclusions}
\label{conclu}
A key barrier to adopting micro-level reserving is its significant methodological difference from macro-level models, such as the Chain-Ladder, and the effort required to develop it. The transition from macro to micro-level models is challenging for insurance companies and difficult for regulators to validate, as the foundational principles differ greatly from those of the Chain-Ladder and development factors. This gap between the two reserving methodologies hinders the adoption of micro-level models in the insurance sector.

To face this issue, this paper introduces a novel perspective to claim reserving by applying principles from population sampling theory, specifically through the use of the augmented inverse probability weighting (AIPW) estimator. The key theoretical contribution of this work is that the AIPW estimator integrates the aggregate perspective of macro-level models with the detailed insights of micro-level models, offering a unified statistical framework that enhances both the robustness and accuracy of reserve estimates. This approach effectively bridges the gap between traditional macro-level and micro-level methodologies. Additionally, the sampling framework can provide new theoretical insights into the statistical inference of reserving methods, including those traditionally viewed as ad-hoc procedures, such as the Chain-Ladder method.

Moreover, adopting a population sampling perspective for reserving also offers a new understanding of the reserving process, emphasizing the importance of addressing the sampling bias caused by the partially observed claims. The population sampling literature provides practical and theoretically sound methods  to address the diverse challenges posed by the sampling bias. Specifically, the use of augmentation terms for correction, weighted balance properties,  weighted estimation equation for model estimation, and generation of synthetic data sets have the potential to improve reserve estimation for a given reserving model. All of this is corroborated empirically

The numerical results are promising, demonstrating that the AIPW estimator, along with other sampling-based estimators, improves the precision of reserve estimates and better captures the variability and uncertainty inherent in claims data, with the AIPW estimator showing potential to outperform traditional models in terms of both accuracy and adaptability.

Future research could explore other methods that address sampling bias and their applications to reserving problems. For example, Equation (\ref{IPWform}) can be used as an alternative to combine reserving models for the reported claims along with the IPW principle. Similarly,  other techniques designed for missing data or those from causal inference contexts may offer valuable insights. Specifically, the process of adjusting model predictions based on observed errors in reported claims aligns with Bayesian approaches using posterior distributions. Thus, incorporating Bayesian models into claim reserving may provide an interesting and valuable perspective. Similarly, this paper demonstrates the benefits of employing sampling-based methods rather than focusing solely on finding the optimal configuration. Future research could further explore the best structures for AIPW estimators, as well as for other methods to have optimal reserving predictions.
 
\section*{Acknowledgments}
This work was partly supported by Natural Sciences and Engineering Research Council of Canada [RGPIN 284246, [RGPIN-2023-04326], and the Canadian Institute of Actuaries [CS000269]. 

\section*{Conflicts of interest or Competing interests }
The authors declare no conflicts of interest or competing interests in this paper, with no financial or personal affiliations that could compromise the objectivity or integrity of the presented work.

\section*{Data Availability}
The data in this study is proprietary data from an insurance company that we are unable to share.

\bibliographystyle{apalike}
\bibliography{references}

\appendix

\end{document}